\documentclass[reprint,amsmath,amssymb,aps,prl]{revtex4-1}
\usepackage{graphicx}
\usepackage{psfrag}

\usepackage{bm}
\begin{document}

\title{Dynamics of earthquake nucleation process represented by the Burridge-Knopoff model}

\author{Yushi Ueda}
%\email{kawamura@ess.sci.osaka-u.ac.jp}

\author{Shouji Morimoto}
\author{Shingo Kakui}
\author{Takumi Yamamoto}
\author{Hikaru Kawamura}
\email{kawamura@ess.sci.osaka-u.ac.jp}
%\altaffiliation{Present Address:
%Department of Earth and Space Science, Faculty of Science,
%Osaka University, Toyonaka, Osaka 560-0043, Japan}
%\affiliation{Department of Earth and Space Science, Faculty of Science,
%Osaka University, Toyonaka, Osaka 560-0043, Japan}

\affiliation{Graduate School of Science, Osaka University,
Toyonaka, Osaka 560-0043, Japan}

\date{\today}

\begin{abstract}
Dynamics of earthquake nucleation process is studied on the basis of the one-dimensional Burridge-Knopoff (BK) model obeying the rate- and state-dependent friction (RSF) law. We investigate the properties of the model at each stage of the nucleation process, including the quasi-static initial phase, the unstable acceleration phase and the high-speed rupture phase or a mainshock. Two kinds of nucleation lengths $L_{sc}$ and $L_c$ are identified and investigated. The nucleation length $L_{sc}$ and the initial phase exist only for a weak frictional instability regime, while the nucleation length $L_c$ and the acceleration phase exist for both weak and strong instability regimes. Both $L_{sc}$ and $L_c$ are found to be determined by the model parameters, the frictional weakening parameter and the elastic stiffness parameter, hardly dependent on the size of an ensuing mainshock. The sliding velocity is extremely slow in the initial phase up to $L_{sc}$, of order the pulling speed of the plate, while it reaches a detectable level at a certain stage of the acceleration phase.  The continuum limits of the results are discussed. The continuum limit of the BK model lies in the weak frictional instability regime so that a mature homogeneous fault under the RSF law always accompanies the quasi-static nucleation process. Duration times of each stage of the nucleation process are examined. The relation to the elastic continuum model and implications to real seismicity are discussed.
\end{abstract}

\maketitle

%%%%%%%%%%%%%%%%%
% INTRODUCTION %
%%%%%%%%%%%%%%%%%

\section{I. Introduction}

 Owing to the recent development of the GPS technology, it has been recognized now that general forms of seismic activity could be of a rich variety, often including various types of slow slip events, {\it e.g.\/}, a preslip, an afterslip, a slow earthquake, {\it etc\/}. From the viewpoint of earthquake forecast, a preslip, {\it i.e.\/}, a slow-slip event occurring prior to a mainshock, would be of special significance. Such a preslip prior to a mainshock is usually associated with the nucleation process which occurs preceeding the high-speed rupture of the mainshock. Then, a wide-spread expectation is that a large earthquake might be preceded by a precursory nucleation process which occurs prior to the high-speed rupture of the mainshock. Such a precursory phenomenon preceding mainshocks, if any, would be of paramount importance in its own right as well as in its possible connection to earthquake forecast.

 Although the nucleation phenomena preceding the main rupture have been more or less confirmed by laboratory rock experiments \cite{Latour,McLasky}, its nature, or even its very existence, remains less clear for real earthquakes \cite{Ando,Ohta,Kato,Bouchon,Tape}. We note that a similar nucleation process is ubiquitously observed in various types of failure processes in material science and in engineering.  

 Nucleation process is supposed to be localized to a compact ``seed'' area with its rupture velocity orders of magnitude lower than the seismic wave velocity \cite{Dieterich92,Ohnaka00,Ohnaka03,Scholzbook,Dieterich09}. The fault spends a very long time in this nucleation process, and then at some point, exhibits a rapid acceleration process accompanied by a rapid expansion of the rupture zone, finally getting into the final high-speed rupture of a mainshock. 

 The earthquake nucleation process might proceed via several distinct steps or ``phases''. According to Ohnaka, it starts with an initial quasi-static process, and gets into the acceleration phase  when the nucleus diameter $L$ exceeds a nucleation length $L_{sc}$, where the system gets out of equilibrium, and rapidly increases its slip velocity \cite{Ohnaka00,Ohnaka03}. Then, when the nucleus diameter exceeds another nucleation length $L_{c} (> L_{sc})$, the fault eventually exhibits a high-speed rupture of a mainshock. In this picture, there appear two characteristic length scales for the nucleus, $L_{sc}$ and $L_{c}$. These two nucleation lengths divide the nucleation process into ``the initial phase''  in which the nucleus size $L$ is smaller the $L_{sc}$ ($L<L_{sc}$), ``the acceleration phase'' in which the nucleus size exceeds $L_{sc}$ but is still smaller than $L_c$ ($L_{sc}<L<L_c$), and ``the high-speed rupture phase'' of mainshock ($L>L_c$). 

 Because of a slow character of the slip, earthquake nucleation process might also be regarded as a type of more general slow-slip phenomena, including afterslips and slow earthquakes. While the relation between these different types of slow seismic processes poses an interesting and important question, we focus in the present paper on the nucleation process {\it prior to\/} the high-speed rupture of mainshock.

 Under such circumstances, in order to get deeper understanding of the physical process behind the seismic nucleation process and the subsequent mainshock, a theoretical or a numerical study based on an appropriate model of an earthquake fault would be important and helpful. In such a physical modeling, the friction force is a crucially important part. The friction force now standard in seismology is the so-called rate and state dependent friction (RSF) law \cite{Dieterich79,Ruina,Marone,Scholz98}. In the pioneering study, Dieterich derived a formula describing the nucleation length based on such RSF law \cite{Dieterich92}. The most standard form of the nucleation length based on the RSF law might be
\begin{equation}
\eta \frac{G\mathcal{L}}{\sigma_n(B-A)},
\label{nucleationlength}
\end{equation}
where $\sigma_n$ is the normal stress, $G$ is the rigidity, $\mathcal{L}$ is a characteristic slip distance and $\eta$ is a constant, while $A$ and $B$ are the frictional parameters associated with the RSF law, each representing the velocity-strengthening and the frictional-weakening parts of the friction. Dieterich also suggested that under certain conditions the nucleation length might be given by \cite{Dieterich92}
\begin{equation}
\eta \frac{G\mathcal{L}}{\sigma_n B},
\label{nucleationlength2}
\end{equation}
in which the $A$ parameter did not appear.

 The RSF law has been used in many of numerical simulations on earthquakes, mostly in the continuum model \cite{TseRice,Stuart,Horowitz,Rice,Ben-ZionRice,KatoHirasawa,Bizzarri06a,Bizzarri06b}, including the earthquake nucleation process. In particular, Ampuero and Rubin studied the properties of the nucleation process for the continuum model under the RSF law, with two representative evolution laws, {\it i.e.\/}, the aging law \cite{AmpueroRubin} and the slip law \cite{RubinAmpuero} within the quasi-static approximation neglecting the inertia effect.

 Meanwhile, a further simplified {\it discrete\/} model has also been used in earthquake studies. Particularly popular is the so-called spring-block model or the Burridge-Knopoff (BK) model \cite{BK}, in which an earthquake fault is modeled as an assembly of blocks mutually connected via elastic springs which are subject to the friction force and are slowly driven by an external force mimicking the plate drive.

 The model might be better justified in the situation where there exists a well-developed mature fault layer, presumably corresponding to the low-velocity fault zone \cite{Huang} observed in many mature faults \cite{Ueda}. The fault layer is supposed to be uniformly pulled by the more or less rigid crust contingent to it. Because of its simplicity, the BK model is particularly suited to the study of statistical properties of earthquakes, since it often enables one to generate sufficiently many events, say,  hundreds of thousands of events, to reliably evaluate its statistical properties. This type of model might also be relevant to the description of other stick-slip-type phenomena such as landslides \cite{Viesca}.

 In many numerical simulations of the BK model, while a simple velocity-weakening friction law in which the friction force is assumed to be a single-valued decreasing function of the velocity has often been used \cite{CL89a,CL89b,CLST,Carlson91,Carlson91b,CLS-review,Schmittbuhl,MoriKawamura05,MoriKawamura06,MoriKawamura08a,MoriKawamura08b,MoriKawamura08c,Kawamura-review}, a more realistic RSF law was also employed in some of recent numerical simulations of the model. For example,  Cao and Aki performed a numerical simulation by combining the one-dimensional (1D) BK model with the RSF law in which various constitutive parameters were set nonuniform over blocks \cite{CaoAki}. Ohmura and Kawamura extended an earlier calculation by Cao and Aki to study the statistical properties of the 1D BK model combined with the RSF law with uniform constitutive parameters \cite{OhmuraKawamura,Kawamura-review}. Clancy and Corcoran also performed a simulation of the model based on a modified version of the RSF law \cite{Clancy09}.

 Of course, the space discretization in the form of blocks is a crude approximation to the original continuum crust. It introduces the short-length cut-off scale into the problem in the form of the block size, which could in principle give rise to an artificial effect not realized in the continuum. Indeed, such a criticism against the BK model was made in the past \cite{Rice}. Rice criticized that the discrete BK model with the simple velocity-weakening law was ``intrinsically discrete'', lacking  in a well-defined continuum limit, arguing that the spatiotemporal complexity observed in the discrete BK model was due to an inherent discreteness of the model, which should disappear in continuum \cite{Rice}.  In contrast to the simple velocity-weakening law, the RSF law possesses an intrinsic length scale corresponding the characteristic slip distance ${\mathcal L}$. If the grid spacing $d$ was taken smaller than the nucleation length which is proportional to this characteristic slip distance ${\mathcal L}$, the system tended to exhibit a quasi-periodic recurrence of large events, whereas, if the grid spacing $d$ was taken larger than it, the system exhibited an apparently complex or critical behavior. Note that the extent of the discreteness may be regarded as a measure of the underlying spatial inhomogeneity \cite{Rice}.

 This problem of the continuum limit of the BK model was also addressed within the velocity-weakening friction law by Myers and Langer \cite{Myers93}, by Shaw  \cite{Shaw94}, and by Mori and Kawamura \cite{MoriKawamura08c}, where the Kelvin viscosity term was introduced to produce a small length scale allowing for a sensible continuum limit. 

 In fact, the problem of the small length scale of the BK model is closely related to the nucleation phenomena. According to Rice, the continuum system under the RSF law always exhibits a quasi-static nucleation process prior to a mainshock \cite{Rice}. In view of such a claim, and also of the general importance of the seismic nucleation phenomena, in the present paper we wish to clarify by means of extensive numerical simulations on the 1D BK model how the nucleation process of the discrete BK model behaves in its continuum limit, by systematically varying the extent of the discreteness of the model. By so doing, we also wish to clarify the nature of the nucleation process of the model, and discuss its implications to real seismicity.

 Of course, since the real fault plane is more or less two-dimensional (2D), the assumed one-dimensinality of our present model is a high simplification. While earlier studies suggested that most of qualitative features of the mainshock statistical properties could be captured even by the 1D model, there exists a possibility that, concerning the nucleation properties, the 2D model might exhibit a behavior different from the 1D model due to the richness of the underlying geometry \cite{Kawamura-review}. In the present paper, leaving a systematic study of the nucleation process of the 2D BK model as a future task, we first study the 1D model which is far simpler than the corresponding 2D model.

 In such a way, on the basis of the 1D BK model, we wish to shed light on the nucleation process of a mature fault, {\it e.g.\/}, the nucleation dynamics, the nucleation lengths and the duration times of each phase of the nucleation process. How these quantities depend on material parameters, and are related or unrelated to the size of the ensuing mainshock ? Such an issue would be of special significance from the standpoint of utilizing earthquake nucleation phenomena for a possible earthquake forecast. For example, if the nucleation length $L_{sc}$ or $L_c$ is correlated with the mainshock size, {\it e.g.\/}, a larger earthquake for a larger $L_{sc}$ or $L_c$, one might have a chance to predict the size of the mainshock from the measurement of the nucleation lengths. If, on the other hand, the nucleation length $L_{sc}$ or $L_c$ is not correlated with the mainshock size, the prediction of the mainshock size from the measurement of the nucleation lengths would be impossible.

 By its nature, the fault sliding velocity in the nucleation process tends to be very low. Hence, for any practical detection, it would crucially be important to clarify how fast the fault sliding velocity is, and how much time is left before the ensuing mainshock. With these motivations in mind, we try to conduct a systematic numerical and analytic study of the BK model in the following part of the paper.

 A preliminary account of our simulations was already given in  \cite{Ueda}. In the present paper, we conduct a systematic survey of more general parameter space, give detailed analytical treatments, and make comparison between theoretical and numerical results.

 The rest of the paper is organized as follows. In section II, we define our model, the 1D BK model obeying the RSF law, and present its equation of motion. Its continuum limit is also given. In section III, we report on the results of our numerical simulations on the dynamics of the nucleation process of the model.  In subsection III[A], we first illustrate mains features of the nucleation process. Two distinct parameter regimes exist, {\it i.e.\/}, the weak frictional instability regime and the strong frictional instability regime. The two kinds of nucleation lengths $L_{sc}$ and $L_c$ are identified. In the subsequent subsections [B]-[D], we present our numerical data on the dynamics of the model at each stage of its nucleation process, {\it i.e.\/}, [B] the initial phase of the weak frictional instability regime, [C] the acceleration phase of the weak frictional instability regime, and [D] the acceleration phase of the strong frictional instability regime. In section IV, we report on the results of our theoretical analyses of the nucleation process of the model. After explaining in subsection [A] the basic scheme of the perturbation method employed, we examine the dynamics of the model in some detail in the following subsections [B]-[D], {\it i.e.\/}, [B] the initial phase, [C] the acceleration phase at which the epicenter-block sliding velocity $v$ is smaller than the crossover velocity $v^*$, $v<v^*$, and [D] the acceleration phase at $v>v^*$.  Analytic expressions of the nucleation length $L_{sc}$ and of the condition discriminating the weak and the strong frictional instability regimes is derived in subsection [B]. In subsection [E], we perform a mechanical stability analysis to re-derive $L_{sc}$ and the weak/strong instability condition, which confirms the results from the perturbation analysis. In section V, we present the results of our numerical simulations focusing on various statistical properties characterizing the nucleation process, including the nucleation lengths $L_{sc}$ and $L_c$, and the duration times of each phase, averaged over many events. Their continuum limits are also examined. Finally, section VI is devoted to summary and discussion. Implications to real seismicity and possible extensions of the present analysis are discussed.

\section{II. The model and its continuum limit}

The 1D BK model consists of a 1D array of $N$ identical blocks of the mass $m$, which are mutually connected with the two neighboring blocks via the elastic springs of the spring stiffness $k_c$, also connected to the moving plate via the springs of the spring stiffness $k_p$, and are driven with a constant rate $\nu'$. All blocks are subject to the friction force $\Phi$, which is the source of the nonlinearity in the model.

 The equation of motion for the $i$-th block can be written as
\begin{equation}
m \frac{{\rm d}^2U_i}{{\rm d}t^{\prime 2}} = k_p (\nu ' t'-U_i) + k_c (U_{i+1}-2U_i+U_{i-1})-\Phi_i,
\label{original-eq-motion}
\end{equation}
where $t'$ is the time, $U_i$ is the displacement of the $i$-th block, and $\Phi_i$ is the friction force at the $i$-th block. For simplicity, the  motion in the direction opposite to the plate drive is inhibited by imposing an infinitely large friction for $\dot U_i<0$.

For the friction law, we assume the RSF law given  by 
\begin{eqnarray}
\Phi_i=\left\{C + A \log(1+\frac{V_i}{V^*}) + B \log \frac{V^* \Theta_i}{{\mathcal L}} \right\} {\mathcal N} ,
\label{original-RSF}
\end{eqnarray}
where $V_i=\frac{{\rm d}U_i}{{\rm d}t^\prime}$ is the sliding velocity of the $i$-th block, $\Theta_i(t')$ is the time-dependent state variable (with the dimension of the time) representing the ``state'' of the slip interface, $V^*$ is a crossover velocity underlying the RSF law, ${\mathcal N}$ is an effective normal load, ${\mathcal L}$ is a critical slip distance which is a measure of the sliding distance necessary for the surface to evolve to a new state, with $A,\ B$ and $C$ positive constants describing the RSF law. The first term ($C$-term) is a constant taking a value around $\frac{2}{3}$, which dominates the total friction in magnitude, the second term ($A$-term) a velocity-strengthening direct term describing the part of the friction responding immediately to the velocity change, the third part ($B$-term) an indirect frictional-weakening term dependent of the state variable $\Theta$. Laboratory experiments suggest that the $A$- and $B$-terms are smaller than the $C$-term by one or two orders of magnitudes, yet they play an essential role in stick-slip dynamics \cite{Marone,Scholz98,Scholzbook}. 

 Note that, in the standard RSF law, the $A$-term is often assumed to be proportional to $\log(\frac{V}{V^*})$. Obviously, this form becomes pathological in the $V\rightarrow 0$ limit because it gives a negatively divergent friction. In other words, the pure logarithmic form of the $A$-term cannot describe the state at rest. We cure this pathology by phenomenologically introducing a modified form given above. The modified form, where the $A$-term becomes proportional to the block velocity $V$ at $V<<V^*$ but reduces to the purely logarithmic form at $V>>V^*$, enables one to describe a complete halt. The characteristic velocity $V^*$ represents a crossover velocity, describing the low-velocity cutoff of the logarithmic behavior of the friction.
 
%  Anyway, the existence or nonexistence of this unity in the logarithm of the $A$-term affects the behavior of the system  against a tectonic loading during the long interseismic period. With this unity in the $A$-term, the response of the system during the interseismic period is a complete stick or a halt with $v=0$, while without this unity the response of the system is a slow steady-state movement without any complete halt. Naturally, either existence or nonexistence of the $A$-term also affects the nucleation process. Here, we suppose that a complete stick with $v=0$ occurs during the interseismic period and assume the $A$-term with unity in the logarithm.					

 For the evolution law of the state variable $\Theta$, we use here the so-called aging (slowness) law given by
\begin{eqnarray}
\frac{d\Theta_i}{dt'}=1-\frac{V_i\Theta_i}{{\mathcal L}}.
\label{original-aging}
\end{eqnarray}
Under this evolution law, the state variable $\Theta_i$ grows linearly with the time at a complete halt $V_i=0$, reaching a very large value at the outset of the nucleation process, while it decays very rapidly during the seismic rupture.

 The equation of motion can be made dimensionless by taking the length unit to be the critical slip distance ${\mathcal L}$, the time unit to be $\omega^{-1}=\sqrt{m/k_p}$ and the velocity unit to be $\mathcal{L}\omega$,
\begin{eqnarray}
\frac{{\rm d}^2u_i}{{\rm d}t^2} &=& \nu t-u_i+l^2(u_{i+1}-2u_i+u_{i-1}) \nonumber \\ 
&-& \left( c+a\log \left(1+\frac{v_i}{v^*}\right)+b\log \theta_i \right) ,
\label{eq-motion} \\
\frac{{\rm d}\theta_i}{{\rm d}t} &=& 1-v_i\theta_i ,
\label{aging}
\end{eqnarray}
where the dimensionless variables are defined by $t=\omega t'$, $u_i=U_i/{\mathcal L}$,  $v_i=V_i/(\mathcal{L}\omega)$, $v^*=V^*/(\mathcal{L}\omega)$, $\nu=\nu'/(\mathcal{L}\omega)$,  $\theta_i=\Theta_i \omega$, $a=A{\mathcal N}/(k_p{\mathcal L})$, $b=B{\mathcal N}/(k_p{\mathcal L})$, $c=C{\mathcal N}/(k_p{\mathcal L})$, while $l \equiv \sqrt{k_c/k_p}$ is the dimensionless elastic stiffness parameter. 

 It is sometimes more convenient to rewrite the equation of motion in terms of the velocity variable $v_i$ instead of the displacement $u_i$. By differentiating (\ref{eq-motion}) with respect to $t$ and by using (\ref{aging}), one gets
\begin{eqnarray}
\frac{d^2v_i}{dt^2} &+&  \frac{a}{v_i+v^*}\frac{dv_i}{dt} - l^2(v_{i+1}-2v_i+v_{i-1})+(1-b) v_i \nonumber \\ 
 &=& \nu - \frac{b}{\theta_i} .
\label{eq-motion2}
\end{eqnarray}
The block displacement $u_i$ can be obtained up to a constant by integrating the velocity $v_i$ with respect to $t$.

 One sees from eqs.(\ref{eq-motion2}) and (\ref{aging}) that the constant frictional parameter $c$ no longer remains in the governing equations, meaning this parameter is essentially irrelevant to the dynamical properties of the model. In our simulations, we use either eq.(\ref{eq-motion}) or (\ref{eq-motion2}) depending on the situation. In solving the high-speed motion, we use eq.(\ref{eq-motion}), while in solving the low-speed motion as realized in the initial phase or the early stage of the acceleration phase, we use eq.(\ref{eq-motion2}).

 The frictional parameter $a$/$b$ tends to suppress/enhance the frictional instability. The earthquake instability is driven primarily by the velocity-weakening $b$-term, while the velocity strengthening $a$-term tends to mitigate the unstable slip toward the aseismic slip. Since the frictional parameters $a$ and $b$ compete in their functions, either $a<b$ or $a>b$ might affect the dynamics significantly. Earthquake properties in this regime of $a>b$ will be reported in a separate paper, with emphasis on the slow-slip phenomena intrinsic to this regime. Meanwhile, we find that the properties of the precursory nucleation process of the model, which occurs preceding a mainshock, do not much depend on the relative magnitude of $a$ and $b$. Although we study in the present paper the nucleation process in the parameter range $a<b$ where the unstable seismic character is dominant in a mainshock, main qualitative features of the nucleation process would not change much even for $a>b$.  

 The setting assumed in the BK model in terms of an earthquake fault embedded in the 3D continuum crust  was examined in \cite{Ueda}. In particular, estimates of typical values of the model parameters for natural earthquake faults are given as $\omega ^{-1}\simeq 1$ [s], ${\mathcal L}\simeq $ a few [cm], and $\frac{{\mathcal N}}{k_p{\mathcal L}}\simeq 10^2-10^3$. This yields $\nu \simeq 10^{-7}-10^{-8}$, $c$ around $10^2-10^3$ while $a$ and $b$ being one or two orders of magnitude smaller than $c$. The crossover velocity $V^*$ and its dimensionless counterpart $v^*$ is hard to estimate though it should be much smaller than unity, and we take it as a parameter in our simulations.

 The continuum limit of the BK model corresponds to making the dimensionless block size $d$, defined by $d=\frac{D}{v_s/\omega}$, to be infinitesimal $d\rightarrow 0$, simultaneously making the system infinitely rigid $l\rightarrow \infty$ with $d=1/l$ \cite{MoriKawamura08c}. The dimensionless distance $x$ between the block $i$ and $i^{\prime}$ is given by

\begin{equation}
x=|i-i^{\prime}|d=\frac{|i-i^{\prime}|}{l} . 
\end{equation}

 As discussed in \cite{MoriKawamura08c}, the 1D equation of motion in the continuum limit is given in the dimensionful form by
\begin{equation}
\frac{{\rm d}^2U}{{\rm d}t'^2} = \omega^2(\nu't'-U) + v_s^2 \frac{{\rm d}^2U}{{\rm d}x^2} - \Phi' ,
\label{continuumeq}
\end{equation}
where $U(x,t')$ is the displacement at the position $x$ and the time $t'$, $\Phi'$ is the friction force per unit mass, while $\omega$ and $v_s$ are the characteristic frequency and  the characteristic wave-velocity ($s$-wave velocity), respectively. Note that the term $-\omega^2U$ representing the plate drive is absent in the standard elasto-dynamic equation.

\section{III. Simulation results I}

In this section, we present the results of our numerical simulations on the dynamical properties of the model. After surveying their main features in subsection [A], we present detailed data in the following subsections separately for each phase of the nucleation process. 

\begin{figure}[ht]
\begin{center}
\includegraphics[scale=0.15]{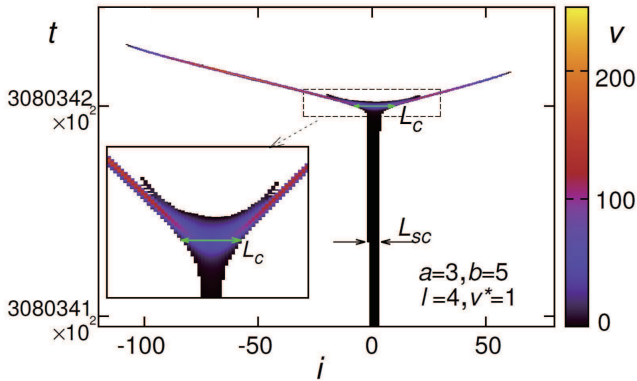}
\end{center}
\caption{
Color plots of typical earthquake nucleation processes depicted in the block-number (position) versus the time plane, realized in the weak frictional instability regime. The color represents the block sliding velocity (white means exactly zero slip, while a low non-zero velocity is represented by black). The model parameters are $a=3$, $b=5$, $c=1000$, $l=4$, $v^*=1$ and $\nu=10^{-8}$. $L_{sc}$ and $L_c$ are two types of nucleation lengths. Events are taken from the event sequence occurring in the stationary state of the model.
}
\end{figure}

\subsection{A. Weak versus strong frictional instability regimes}

 The first question might be whether the 1D BK model under the RSF law ever exhibits a nucleation process prior to a mainshock, and if it does, under what conditions. Remember that our constitutive law allows for a complete stick ({\it i.e.\/}, $v_i=0$ for all $i$) during the interseismic period, which enables us to unambiguously define the onset of the nucleation process by the point where one of the blocks gains a nonzero velocity.

\begin{figure}[ht]
\begin{center}
\includegraphics[scale=0.9]{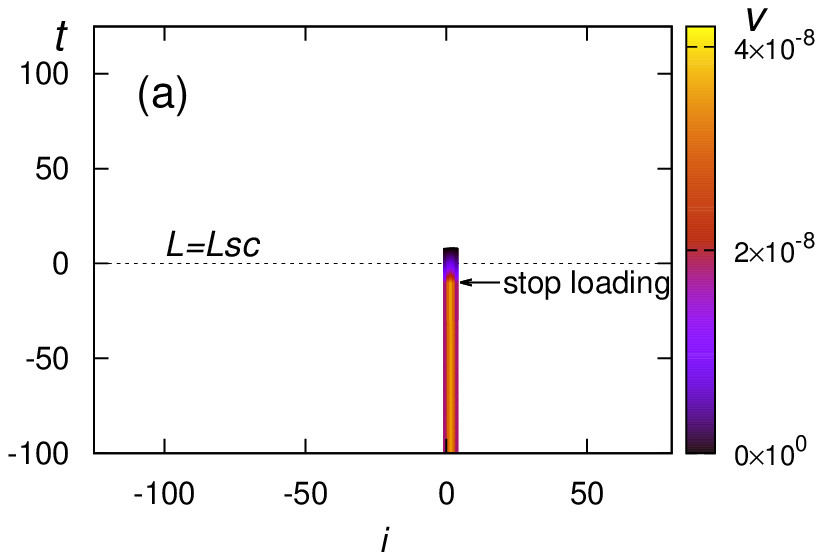}
\includegraphics[scale=0.15]{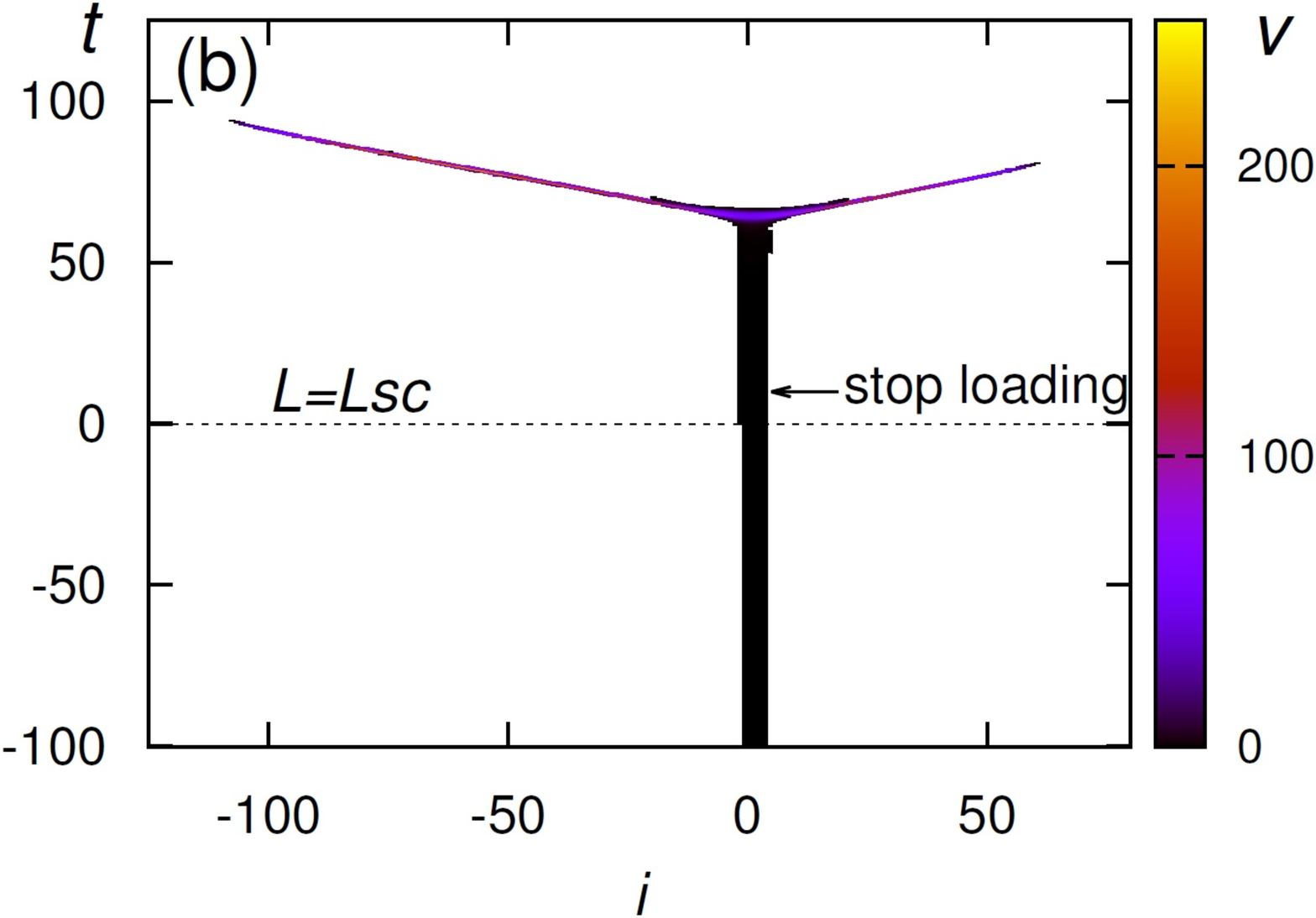}
\end{center}
\caption{
The color plots representing the evolutions of the rupture process when the external loading is artificially stopped, (a) at a point when the number of moving blocks (rupture-zone size) $L$ is less than $L_{sc}$, {\it i.e.\/}, $L=3 < L_{sc}=3.35$, or (b) at a point when the number of moving blocks is greater than $L_{sc}$, {\it i.e.\/}, $L=4 > L_{sc}=3.35$, in the case of the weak frictional instability. The time origin is set to the point of $L=L_{sc}$ here. The color represents the block sliding velocity (white means exactly zero slip). The model parameters are $a=3$, $b=5$, $c=1000$, $l=4$, $v^*=10^{-2}$ and $\nu =10^{-8}$, corresponding to $L_{sc}=3.35$.
}
\end{figure}

 We illustrate in Fig.1 a typical example of seismic events realized in the 1D BK model under the RSF law, where the time evolution of the movement of each block is shown as a color plot. In the figure, the two types of nucleation lengths $L_c$ and $L_{sc}$ are also illustrated. The nucleation process including the initial phase and the accerrelation phase is realized in addition to the high-speed rupture phase (mainshock). The model parameters are set to $a=3$, $b=5$, $c=1000$, $l=4$, $v^*=1$ and $\nu=10^{-8}$. The origin of the time ($t=0$) is taken to be the onset of the nucleation process where an epicenter block begins to move. An example shown in Fig.1 is an event occurring in the stationary state of the seismic sequence of the model, realized after transient initial events where the memory of initial conditions are still remnant.

 As can clearly be seen from the figure, a slow nucleation process with a long duration time of order $t\sim 10^8$ is observed. Among the two types of nucleation lengths, $L_{sc}$ and $L_{c}$ ($L_{sc} < L_c$) in Fig.1, $L_{sc}$ is the length separating stable and unstable ruptures. Namely, when the nucleus size $L$ is less than $L_{sc}$, the rupture process is stable and reversible, whereas, when $L$ exceeds $L_{sc}$, it becomes unstable and irreversible.

 One illustrative way to demonstrate the expected borderline behavior across the nucleation length $L_{sc}$ may be to artificially stop the external loading in the course of a simulation. Indeed, when the external loading is stopped at a point before $L=L_{sc}$, the rupture itself also stops there, as demonstrated in Fig.2(a), whereas, if the external loading is stopped at any point beyond $L=L_{sc}$, the subsequent seismic rupture is no longer stoppable and evolves until its very end, as demonstrated in Fig.2(b). (Even better criterion might be whether the block sliding velocity is increased or decreased when the loading is artificially stopped, rather than whether the block is completely stopped or not.)

 Ohnaka suggested that, in addition to the nucleation length $L_{sc}$, there exists another nucleation length $L_{c}(>L_{sc})$, which discriminates between the acceleration phase and the high-speed rupture phase \cite{Ohnaka00,Ohnaka03}. In the high-speed rupture phase beyond $L_c$, the rupture propagates with a nearly constant speed in both directions in the form of two separate packets of moving blocks, as can be seen in Fig.1. In the figure, the high-speed rupture of a mainshock corresponds to the linear portion of the rupture propagation line with its slope being the propagation speed of $\sim l$. 

 While there might be several ways to define the nucleation length $L_c$ ($> L_{sc}$), we tentatively give one definition here. As can be seen from Fig.1, the number of simultaneously moving blocks $L$ tends to be large around $L_c$. Hence, we tentatively define $L_c$ by the size of the nucleus at which the number of simultaneously moving blocks becomes maximum for a given event, which we denote $L_c^\prime$: See Fig.10(a) below.  At or very close to this point, the epicenter block ceases to move and the group of moving blocks are detached into two parts, each part propagating in opposite directions. Below in subsection D, we shall give another, perhaps physically more appropriate definition of $L_c$, which is actually the one indicated as $L_c$ in Figs.1 and 3.
%{\it i.e.\/}, the $L$-value at which the epicenter-block sliding velocity takes a maximum, will be given below in subsection D below.

 In fact, we find that the slow and reversible nucleaton process corresponding to the initial phase is realized only in the ``weak frictional instability'' regime where the frictional-weakening parameter $b$ is smaller than a critical value $b_c$, while, in the ``strong frictional instability'' regime where the frictional-weakening parameter $b$ is greater than a critical value $b_c$, the nucleation process does not accompany the slow and reversible initial phase. A typical example of seismic events in the strong frictional instability regime is given in Fig.3, where the model parameters are set to $a=1$ and $b=40$, other parameters being common with those of Fig.1. An apparently nucleation-like process seen in Fig.3 just before the high-speed rupture propagation is not a quasi-static initial phase, but is an unstable acceleration phase. Its duration time is around $t\sim 10$ which is by many orders of magnitude shorter than the duration time of the quasi-static initial phase seen in Fig.1, $t\sim 10^8$. The acceleration phase in the stronger frictional instability regime sometimes could be longer, say, $t\sim10^2-10^3$, particularly for smaller $v^*$. Yet, the dynamics is already irreversible there.

\begin{figure}[ht]
\begin{center}
\includegraphics[scale=0.15]{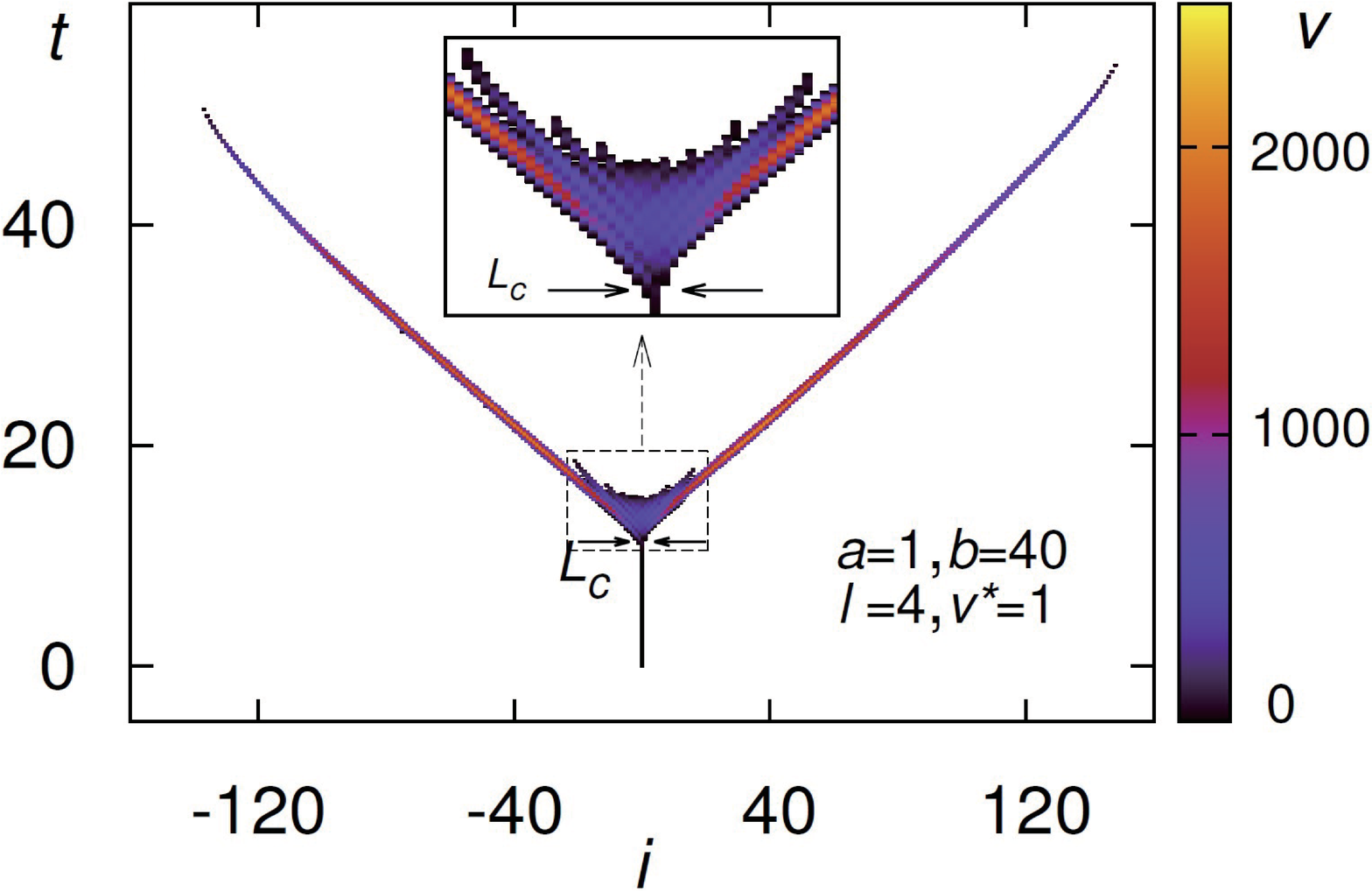}
\end{center}
\caption{
Color plots of typical earthquake nucleation processes depicted in the block-number (position) versus the time plane, realized in the strong frictional instability regime. The color represents the block sliding velocity (white means exactly zero slip). The model parameters are $a=1$, $b=40$, $c=1000$, $l=4$, $v^*=1$ and $\nu=10^{-8}$ ($c$, $l$, $v^*$ and $\nu$ are taken to be common with those in Fig.1). Events are taken from the event sequence occurring in the stationary state of the model.
}
\end{figure}

 The borderline between the weak and the strong frictional instability regimes is given by a critical value of $b$, $b_c$, which is found to have a simple expression $b_c=2l^2+1$. This expression was already reported in Ref.\cite{Ueda}, but we shall derive this expression in several ways in the following part of the paper.

 We note that large events in the weak frictional instability regime always accompany precursory nucleation process irrespective of each individual event, or the choice of the initial conditions, while, those in the strong frictional instability does not accompany a precursory nucleation process in any condition. Hence, for a given set of model parameters, the presence or absence of the quasi-static nucleation process is uniquely determined, not depending on each individual event, but is determined simply by the condition of the friction parameter $b$ being either greater or smaller than the critical value $b_c(l)=2l^2+1$, $l$ being the elastic stiffness parameter.

 In order to demonstrate the spatiotemporal evolution of the nucleation process of the model, we show in Figs.4 and 5 the time evolutions of the spatial profile of (a) the block sliding velocity $v$, (b) the state variable $\theta$, and (c) the multiple of the two quantities $v\theta$, in a typical nucleation process of a large event realized in the stationary state in the weak frictional instability regime. Fig.4 covers the time regime from the onset of the nucleation process till the system reaches $L=L_c$, whereas Fig.5 from the point of $L=L_c$ till an earlier stage of the high-speed rupture phase.  The model parameters are set to $a=1, b=9, l=4, v^*=10^{-2}$ and $\nu=10^{-8}$. From these figures, the manner how the nucleus grows and how the nucleation process transforms into the high-speed rupture of a mainshock is clearly visible.

 Some characteristic points of the nucleation process corresponding to $L=L_{sc}$ and $v=v_{inertia}$ (in Fig.4), $L=L_c$ (in Figs.4 and 5), and $L=L_c^\prime$ (in Fig.5) are indicated  by blue curves. Here $v=v_{inertia}$ is a characteristic crossover velocity at which the inertia effect becomes significant, to be defined below in \S IVD. Note that, beyond the point $v\simeq v_{inertia}$, the inertia effect plays an important role, and the quasi-static approximation in no longer valid. As such, the time range beyond $v\simeq v_{inertia}$ is not covered by \cite{RubinAmpuero} who employed the quasi-static approximation. In the range up to $v\simeq v_{inertia}$, the profiles obtained here look similar to the ones given in \cite{RubinAmpuero} for the continuum model.

\begin{figure}[ht]
\begin{center}
\includegraphics[scale=0.8]{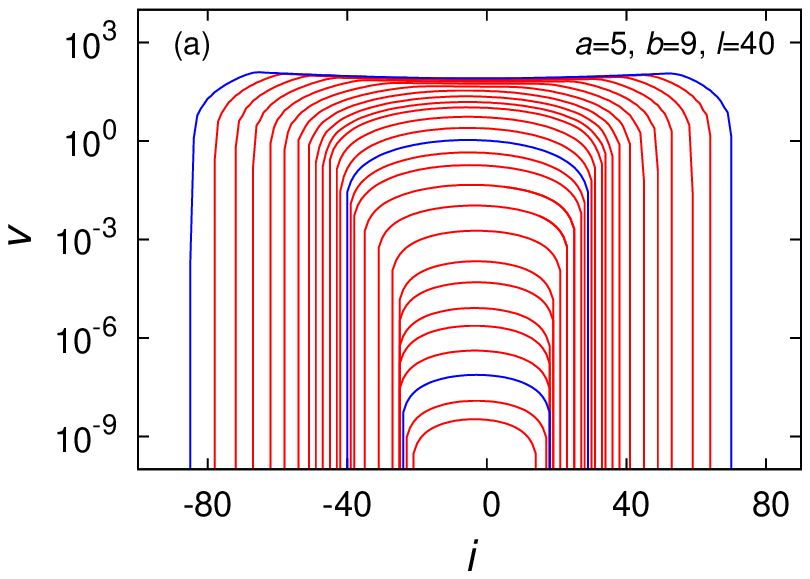}
\includegraphics[scale=0.8]{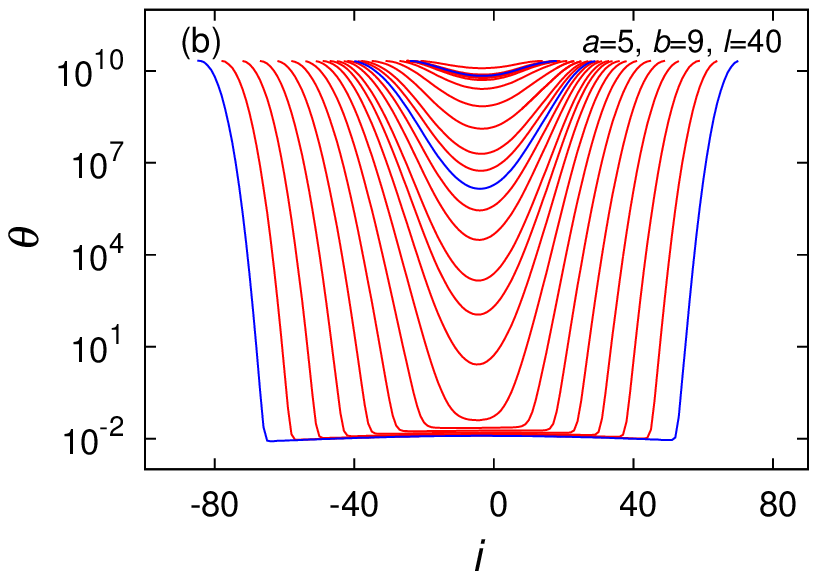}
\includegraphics[scale=0.8]{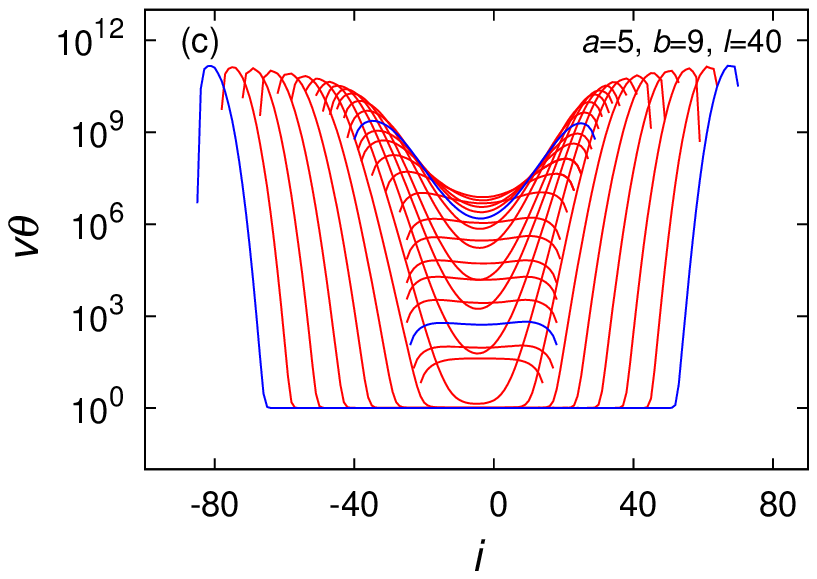}
\end{center}
\caption{
The time evolutions of the spatial profile of (a) the block sliding velocity $v$, (b) the state variable $\theta$, and (c) the multiple of the two quantities $v\theta$, during a typical nucleation process in the weak frictional instability regime.   The model parameters are set to $a=1, b=9, l=4, v^*=10^{-2}$ and $\nu=10^{-8}$. The epicenter block is located at the center, $i=0$. The covered time range is from the onset of the nucleation process until the point of $L=L_c$. The blue curves represent the points of $L=L_{sc}$, $v=v_{inertia}$ and $L=L_c$: See the text for the definitions of $L_{sc}$, $v_{inertia}$ and $L_c$.
}
\end{figure}

 As can be seen from Fig.4(a), the sliding velocity $v$ gets larger until $L=L_c$. Beyond this point, first the epicenter block, and subsequently the neighboring blocks, begin to decelerate, and eventually come to stop (Fig.5(a)). The nucleus is detached into two parts, each of which propagates in the opposite directions forming a rupture front of a mainshock.

 As can be seen from Fig.4(b), in an earlier period of the nucleation process, the state variable $\theta$ maintains its large value acquired during the halt period between mainshocks, while it rapidly decreases in the later period as the block movement accelerates, and eventually reaches a minimum value around $L=L_c$, first at the epicenter block, and subsequently at the neighboring blocks. After this point, the $\theta$-value tends to be recovered again (Fig.5(b)).

 The multiple of $v$ and $\theta$, $v\theta$, plays an important role in the healing process since it appears on the r.h.s. of the equation of motion of the state variable, eq.(\ref{aging}). As can be seen from Figs.4(c) and 5(c), this quantity tends to increase in the earlier period of the nucleation process, first gradually and more rapidly beyond $L=L_{sc}$, reaches a maximum at a point between $L=L_{sc}$ and $L=L_c$, then drops very sharply until it tends to stay around a value close to unity. Note that $v\theta=1$ is a special point corresponding to the stationary condition for the time evolution of the state variable: see eq.(\ref{aging}). Such a plateau-like behavior of $v\theta$ arises around $L_c$ in the epicenter region, and transmits outwards in the nucleus. Further beyond $L_c$, $v\theta$ tends to decrease again, first in the epicenter region, and subsequently in the outer region in the nucleus. % Further details of the block motion and the nucleus expansion during the nucleation process will be given in the following subsections.

\begin{figure}[ht]
\begin{center}
\includegraphics[scale=0.8]{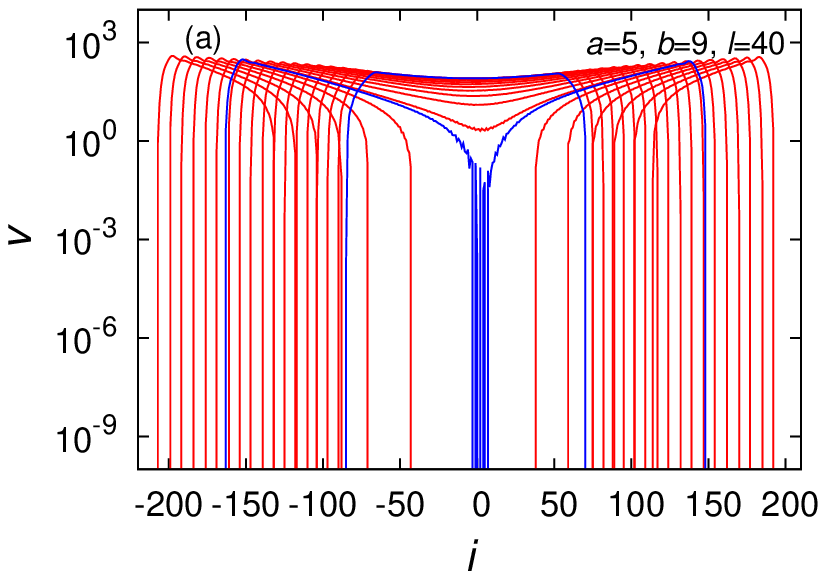}
\includegraphics[scale=0.8]{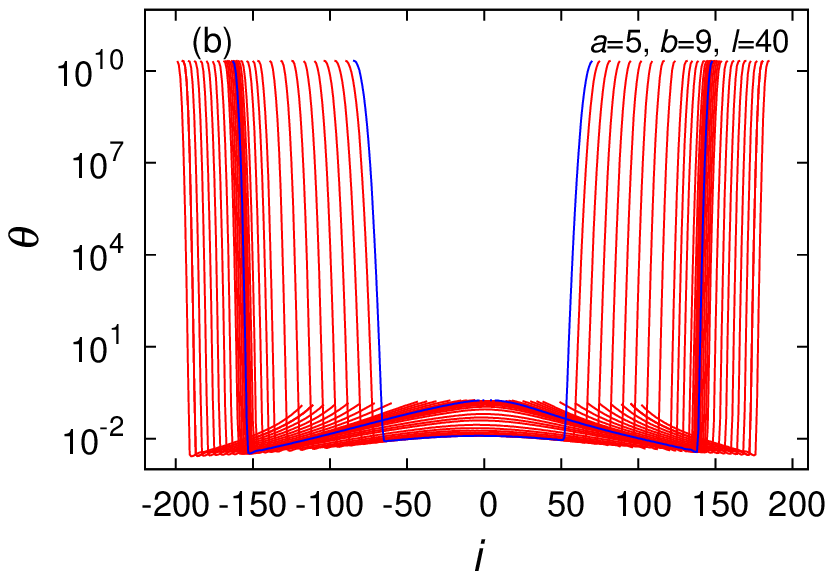}
\includegraphics[scale=0.8]{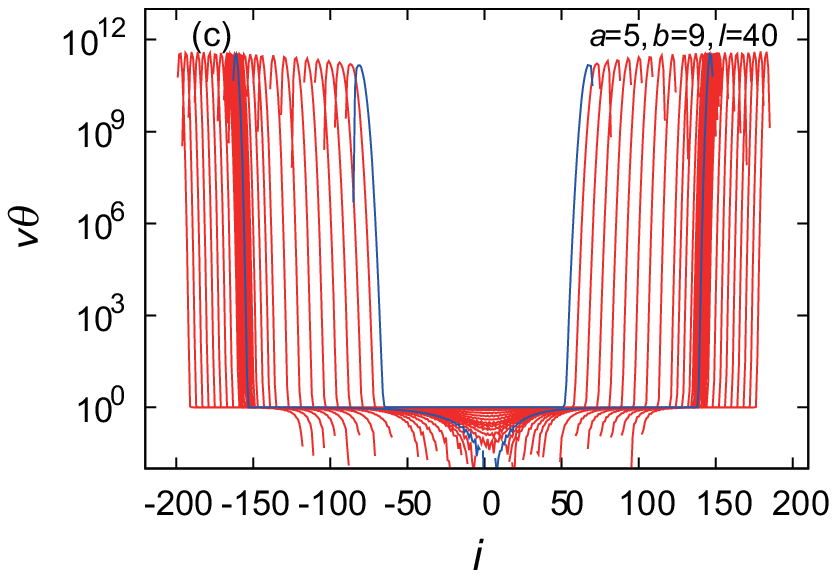}
\end{center}
\caption{
The time evolutions of the spatial profile of (a) the block sliding velocity $v$, (b) the state variable $\theta$, and (c) the multiple of the two quantities $v\theta$, during a typical nucleation process in the weak frictional instability regime.   The model parameters are set to $a=1, b=9, l=4, v^*=10^{-2}$ and $\nu=10^{-8}$.  The epicenter block is located at the center, $i=0$. The covered time range is from the point of $L=L_c$ until the earlier stage of the high-speed rupture of a mainshock, following the time range covered by Fig.4. The blue curves represent the points of $L=L_c$ and $L=L_c^\prime$: See the text for the definitions of $L_c$ and $L_c^\prime$.
}
\end{figure}

 In the following subsections, we present our simulation data in some detail in each phase of the nucleation process, {\it i.e.\/}, (B) the initial phase of the weak frictional instability regime, (C) the acceleration phase of the weak frictional instability regime, and (D) the acceleration phase of the strong frictional instability regime, respectively.

\subsection{B. The initial phase of the weak frictional instability regime}

 Let us begin with the nucleation process in the weak frictional instability regime at $b<b_c=2l^2+1$. 
\begin{figure}[ht]
\begin{center}
\includegraphics[scale=0.73]{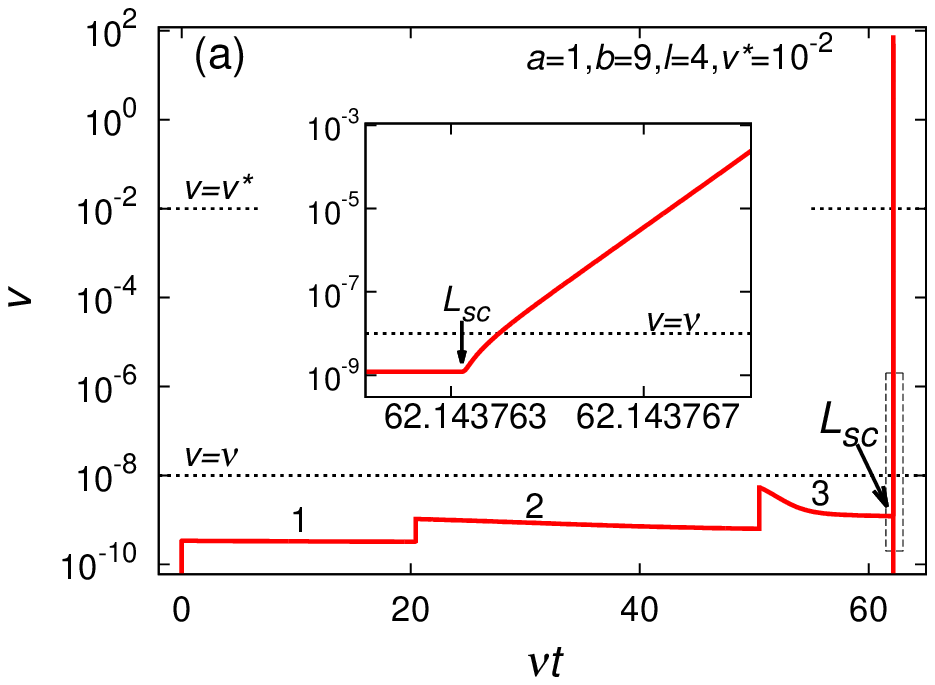}
\includegraphics[scale=0.73]{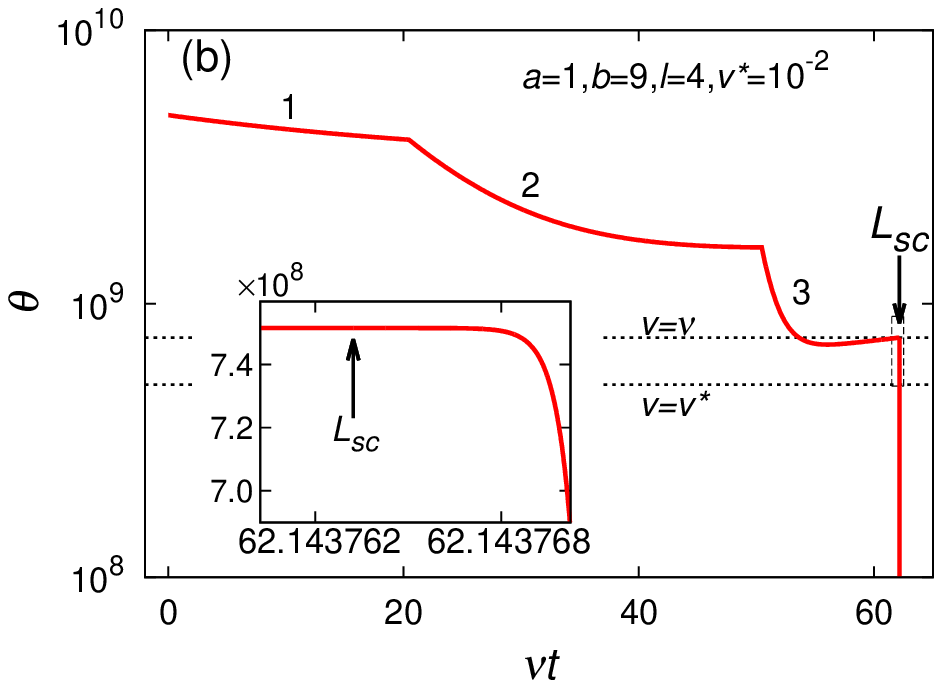}
\includegraphics[scale=0.73]{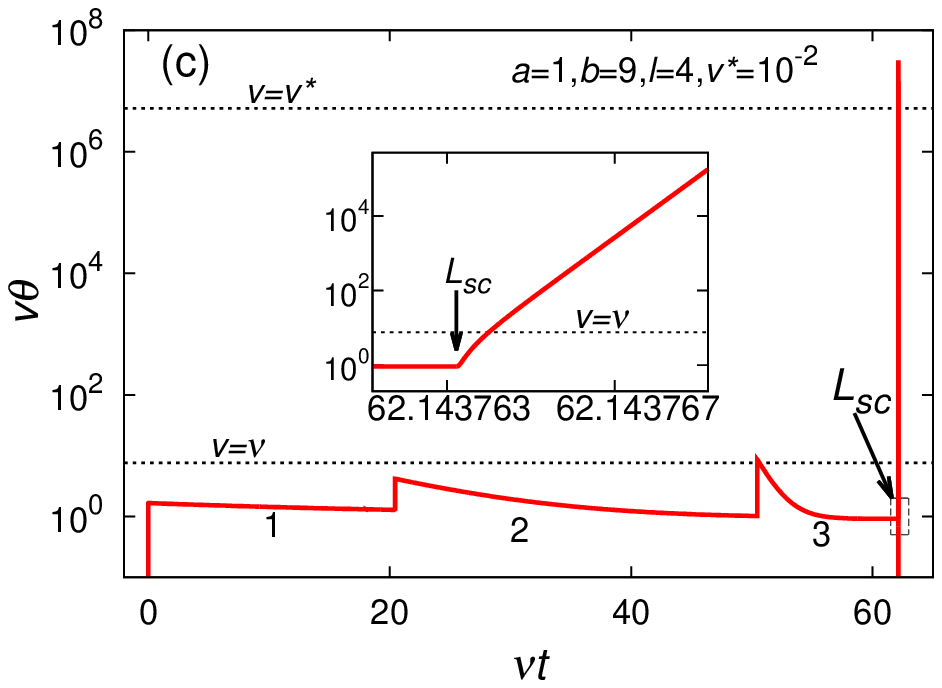}
\end{center}
\caption{
The time evolutions of (a) the sliding velocity $v$, (b) the state variable $\theta$, and (c) the multiple of the two $v\theta$, of an epicenter block in the nucleation process in the weak frictional instability regime. The model parameters are $a=1$, $b=9$, $c=1000$, $l=4$, $v^*=10^{-2}$ and $\nu=10^{-8}$. The time origin $t=0$ is set to the point where the epicenter begins to move. The arrow indicates the point of $L=L_{sc}=3.35$.  The dotted horizontal lines represent the lines corresponding to $v=\nu$, and to $v=v^*$. The integers attached to the curves indicate the number of moving blocks $L$. The insets are magnified views of the region around $L_{sc}$.
}
\end{figure}
 In order to see how the nucleation process evolves with the time, we show in Fig.6 typical time evolutions of (a) the block sliding velocity $v$, (b) the state variable $\theta$, and (c) the multiple of the two quantities $v\theta$, of an epicenter block in a typical nucleation process of a large event realized in the stationary state.  The origin of the time ($t=0$) is set to be the onset of the nucleation process of the event. The model parameters are set to $a=1$, $b=9$, $l=4$,  $v^*=10^{-2}$ and $\nu=10^{-8}$. The inequality $b<b_c=2l^2+1$ is well satisfied, indicating that the system is in the weak frictional instability regime. The nucleation length $L_{sc}$ estimated from eq.(\ref{Lsc}) to be given below is $L_{sc}=3.35$. The discreteness of the model is eminent in this regime.

  At an early stage of the nucleation process, only an epicenter block moves. After some time, the neighboring blocks join this move one by one, causing a spatial expansion of the nucleus. As can be seen from Fig.6(a), the velocity of an epicenter block exhibits a step-like behavior, {\it i.e.\/}, it exhibits an almost discontinuous rise when the block contingent to the moving blocks begins to move joining the nucleation process. As $L_{sc}=3.35$ here, the system gets into the acceleration phase as soon as the number of blocks is increased from 3 to 4, and the epicenter-block sliding velocity begins to increase sharply. The block motion in the subsequent acceleration phase will be examined in the next subsection (Fig.6 to be continued to Fig.8).

 One important general observation is that the epicenter-block sliding velocity  in the initial phase stays very low up to $L=L_{sc}$, of order the pulling speed of the plate $\nu$. This property can also be derived analytically as shown in \S IVA below. In real faults, the plate motion is extremely slow, $\nu \simeq$ a few [cm/year] $\simeq $ 1 [nm/sec].  Real-time detection of such a slow sliding motion would practically be impossible.

 The state variable $\theta$ of an epicenter block initially takes a large value as shown in Fig.6(b). This is simply because $\theta$ linearly increases during the interseismic period according to eq.(\ref{aging}), acquiring a large value just before the onset of the nucleation process. During the initial phase, $\theta$ still keeps its large value since the velocity is still small on the r.h.s. of eq.(\ref{aging}), while it drops steeply beyond $L_{sc}$. The quantity $v\theta$ increases with the time beyond $L_{sc}$, as can be seen from Fig.6(c).

 We note that the dynamics of the model as shown here does not change much depending on the $v^*$-value or on the $a$-value as long as $a$ is taken smaller than $b$, though the time evolution tends to be milder for smaller $v^*$ or larger $a$. This tendency can naturally be understood because the smaller $v^*$ or the larger $a$ in eq.(\ref{eq-motion}) means a larger contribution of the velocity-strengthening $a$-term. The velocity-strengthening force serves to soften an abrupt change, causing a smoother time-evolution of observables.

 The parameter choice of Fig.6 corresponds to $L_{sc}=3.35$ and the discreteness of the model tends to be important around $L_{sc}$. In order to examine the effect of the discreteness on the nucleation dynamics, and to examine an approach to the continuum limit, we show in Fig.7 the corresponding figures for a different set of the parameters corresponding to Figs.4 and 5, {\it i.e.\/},  $a=5$, $b=9$, $l=40$ (other parameters are the same as in Fig.6), which yields a larger $L_{sc}$-value of $L_{sc}=43.42$. One can see, while the nucleation process becomes smoother in this near-continuum case as expected, qualitative features remain similar to those observed in the strongly discrete case of Fig.6.

\begin{figure}[ht]
\begin{center}
\includegraphics[scale=0.73]{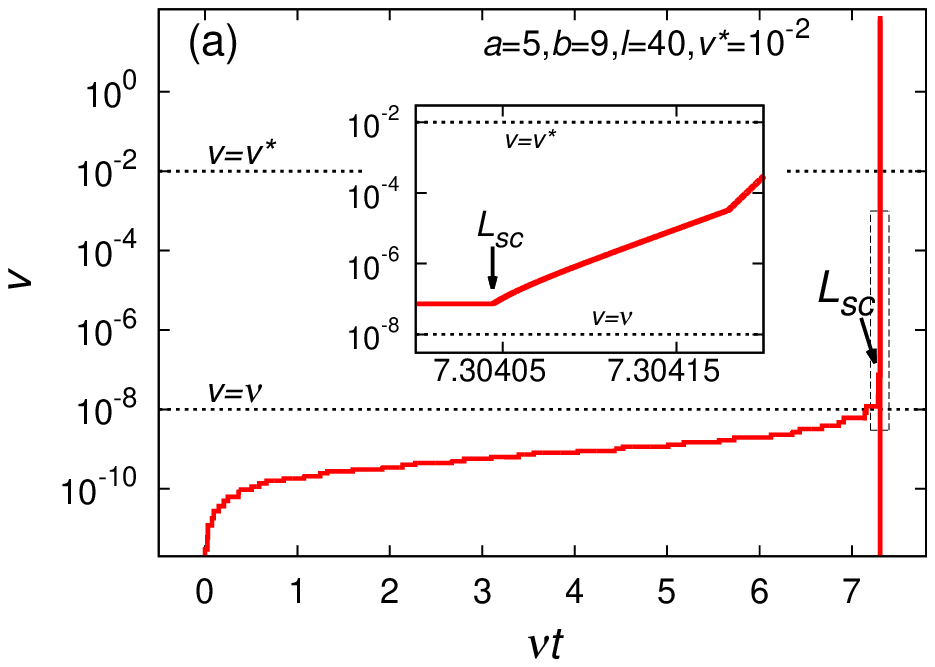}
\includegraphics[scale=0.73]{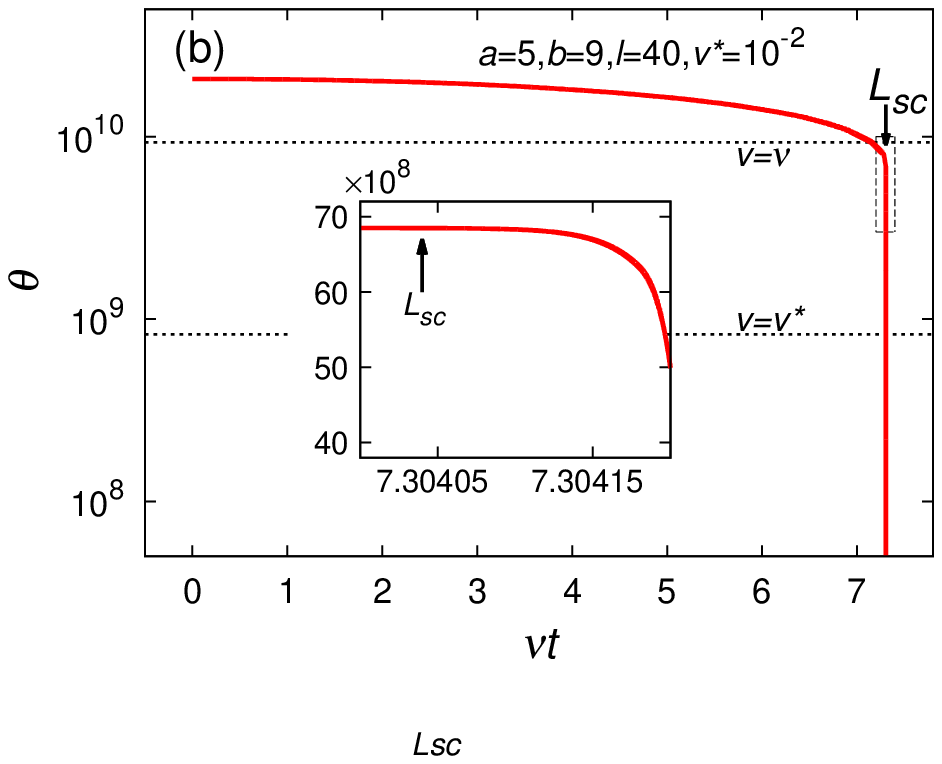}
\includegraphics[scale=0.73]{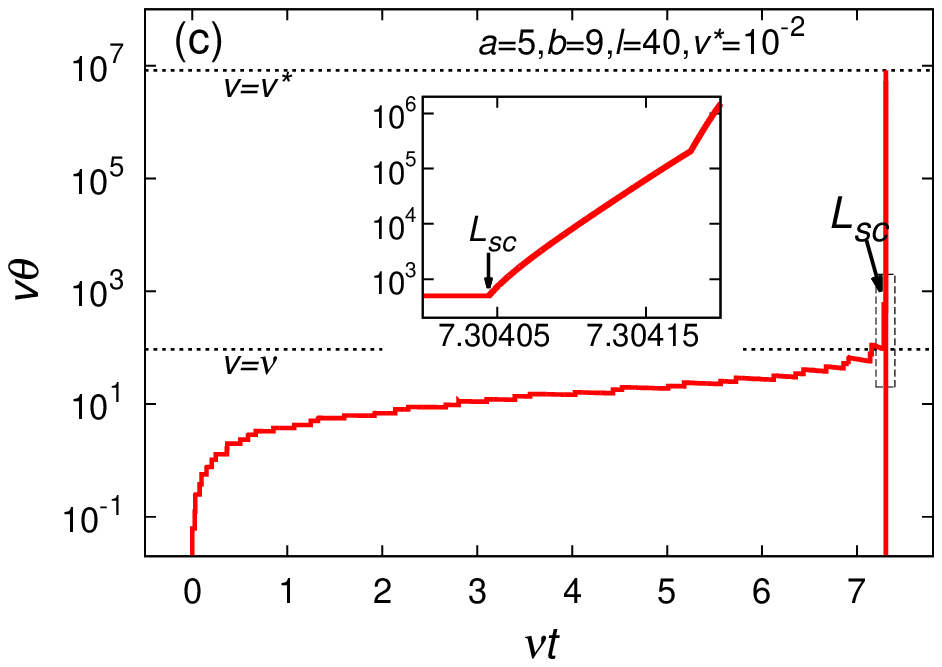}
\end{center}
\caption{
The time evolutions of (a) the sliding velocity $v$, (b) the state variable $\theta$, and (c) the multiple of the two $v\theta$ (c), of an epicenter block in the nucleation process in the weak frictional instability regime. The model parameters are $a=5$, $b=9$, $c=1000$, $l=40$, $v^*=10^{-2}$ and $\nu=10^{-8}$, corresponding to the near-continuum regime. The time origin $t=0$ is taken at the point where the epicenter begins to move. The arrow indicates the point of $L=L_{sc}=43.42$. The dotted horizontal lines represent the lines corresponding to $v=\nu$, and to $v=v^*$. The insets are magnified views of the region around $L_{sc}$. 
}
\end{figure}

 {\it A characteristic feature of the block motion in the quasi-static initial phase is that there exist two different time scales: a slow motion of the time scale $O(1/\nu)$ and a faster one of the time scale $O(1)$\/}. The former might be better described by the slow time variable $\tau\equiv \nu t$. Indeed, a perturbative treatment to be given in \S IVA yields the time evolutions of the sliding velocity $v$ and of the state variable $\theta$ of the epicenter block as
\begin{eqnarray}
v(\tau, t) &=& C_+e^{\lambda_+t}+C_-e^{\lambda_-t} \nonumber \\ 
&+& \frac{1}{\xi_L-b}\left( 1-\frac{b}{\xi_L+(\nu \theta_0-\xi_L)e^{-\frac{\tau}{\xi_L-b}}}\right) ,
\label{v-initial}
\end{eqnarray}
\begin{equation}
\theta (\tau)= \frac{1}{\nu} \left( \xi_L+(\nu \theta_0-\xi_L)e^{-\frac{\tau}{\xi_L-b}} \right),
\end{equation}
where $C_\pm$ are constants to be determined by initial conditions, $\theta_0$ is the $\tau=0$ value of $\theta$, and
\begin{equation}
\lambda_\pm = -\frac{a}{2v^*}\pm \sqrt{ \left(\frac{a}{2v^*}\right)^2 + b-\xi_L } ,
\label{lambda}
\end{equation}
with $\xi_L$ defined by 
\begin{equation}
\xi_L = 2l^2\left(1-\cos\frac{\pi}{L+1} \right)+1 .
\label{xiL}
\end{equation}
In the solution, the number of simultaneously moving blocks (the nucleus size) $L$ is assumed to be fixed during the block movement.

  When the number of moving blocks or the nucleus size $L$ is small such that $b-\xi_L<0$, both $\lambda_+$ and $\lambda_-$ are negative. When the condition $b-\xi_L=0$ is reached, $\lambda_+$ changes its sign, leading to the instability. In fact, this condition $b=\xi_L$ determines the point of $L=L_{sc}$. From eq.(\ref{v-initial}), one can show that the block sliding velocity stays of order $\nu$ throughout the initial phase up to $L=L_{sc}$.

\subsection{C. The acceleration phase of the weak frictional instability regime}

 Next, we proceed to the acceleration phase of the weak frictional instability regime, which occurs beyond $L_{sc}$ succeeding the initial phase. In the acceleration phase, the block movement exhibits a prominent acceleration,  no longer quasi-static nor reversible. 

 In Figs. 8 and 9, we show typical time evolutions of (a) the sliding velocity $v$, (b) the state variable $\theta$, and (c) the multiple of the two $v\theta$, of an epicenter block. The model parameters are taken to be the same as those of Figs.6 and 7, the former corresponding to the strongly discrete case, and Fig.7 to the near-continuum case.  The origin of the time ($t=0$) is set here to the point of $L=L_{sc}$. Note the difference in the time scales from those in Figs. 6 and 7: The abscissa in Figs.8 and 9 is $t$, instead of $\nu t$ in Figs.6 and 7. The arrows in the figures indicate the points of $L=L_{sc}$ and of $L=L_c$, respectively.

\begin{figure}[ht]
\begin{center}
\includegraphics[scale=0.73]{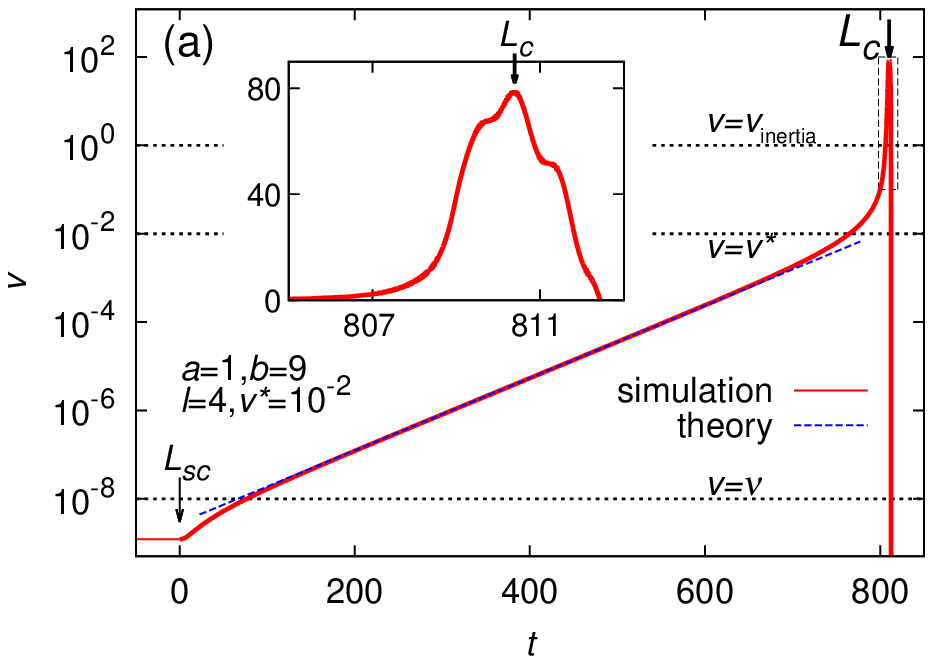}
\includegraphics[scale=0.73]{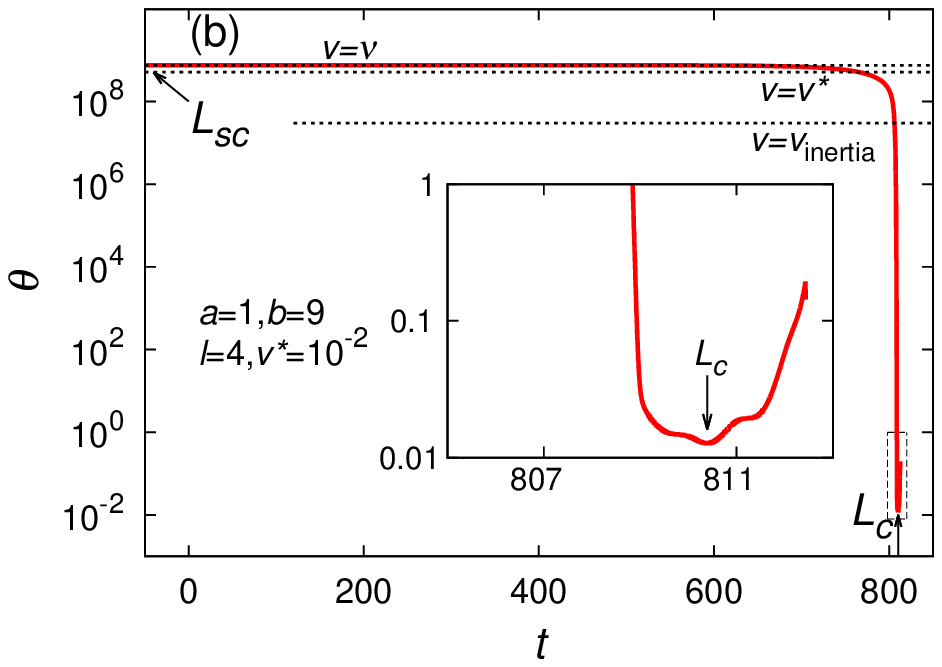}
\includegraphics[scale=0.73]{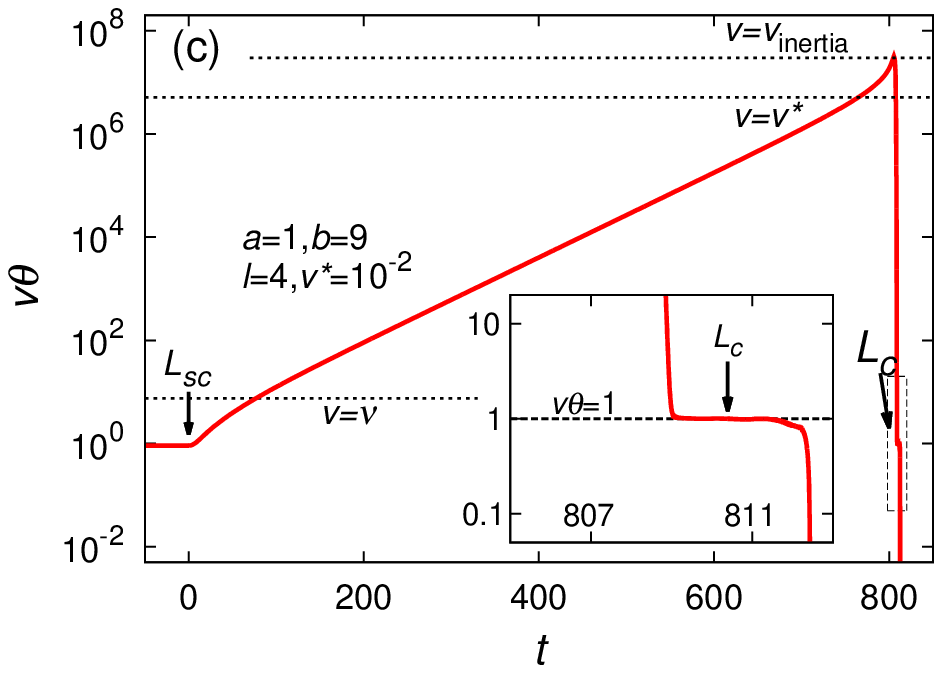}
\end{center}
\caption{
The time evolutions of (a) the sliding velocity $v$, (b) the state variable $\theta$, and (c) the multiple of the two $v\theta$, of an epicenter block in the acceleration phase in the weak frictional instability regime. The model parameters are $a=1$, $b=9$, $c=1000$, $l=4$, $v^*=10^{-2}$ and $\nu=10^{-8}$, corresponding to the strongly discrete case. The time origin $t=0$ is taken at the point of $L=L_{sc}$. The arrows indicate the points of $L=L_{sc}=3.35$ and of $L=L_c=14$. The dotted horizontal lines represent the lines corresponding to $v=\nu$, $v=v^*$ and $v=v_{inertia}$. The blue line in (a) is the theoretical curve for fixed $L$, eq.(\ref{vL}).
}
\end{figure}
%The insets are magnified views of the region around $L_{c}$. 
%(to be defined in subsection C).
%

 As can be seen from Figs. 8(a) and 9(a), the epicenter-block sliding velocity increases rapidly in the acceleration phase, reaching a maximum of order $v\simeq 10^0\sim 10^2$, then decreases sharply and finally stops around $L_c$. The state variable, which stayed nearly constant keeping its large value of order $1/\nu$ throughout the initial phase, begins to drop in the acceleration phase, and eventually becomes of order unity. Since the increase in $v$ dominates over the decrease in $\theta$ at an earlier stage of the acceleration phase, $v\theta$ increases for some period, reaches a maximum, then drops sharply until it becomes close to unity: See Figs.8(c) and 9(c). Note that, around $v\theta =1$, the time variation of $v\theta $ tends to level off exhibiting a much slower time dependence as can be seen from the inset. It is an inevitable consequence of the equation of motion, eq.(\ref{aging}).

\begin{figure}[ht]
\begin{center}
\includegraphics[scale=0.73]{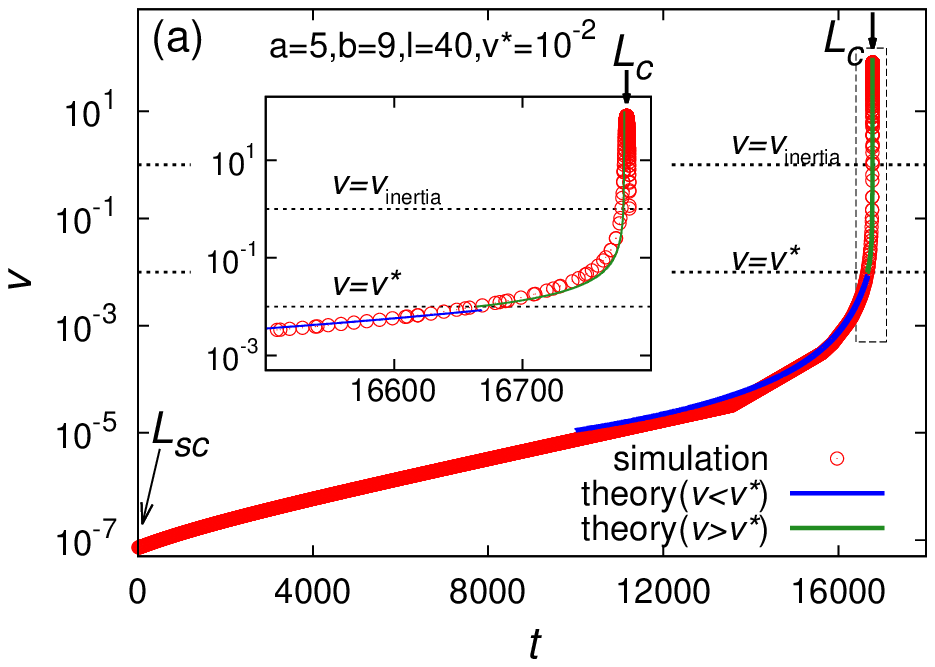}
\includegraphics[scale=0.73]{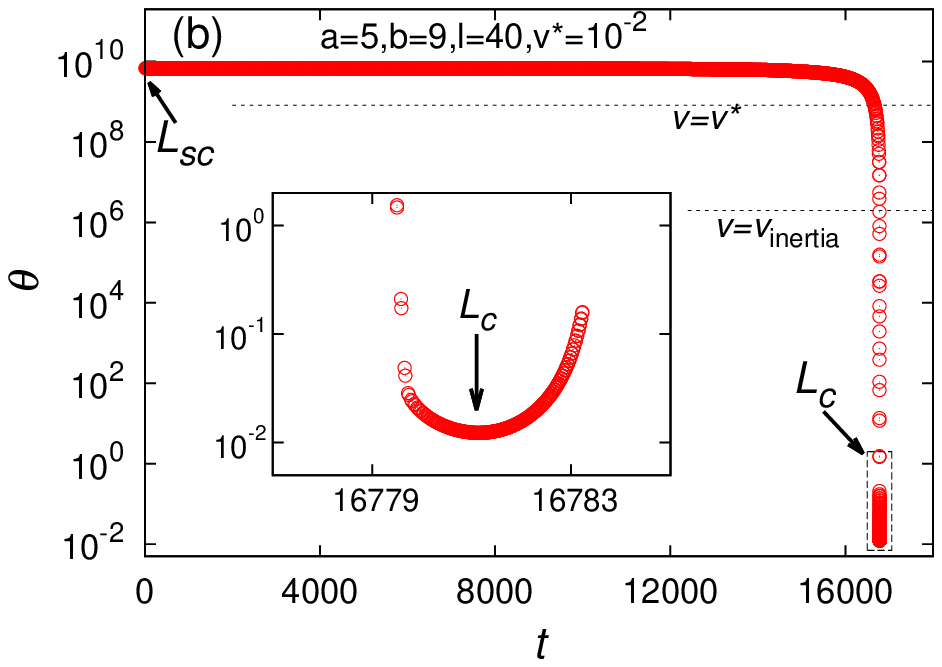}
\includegraphics[scale=0.73]{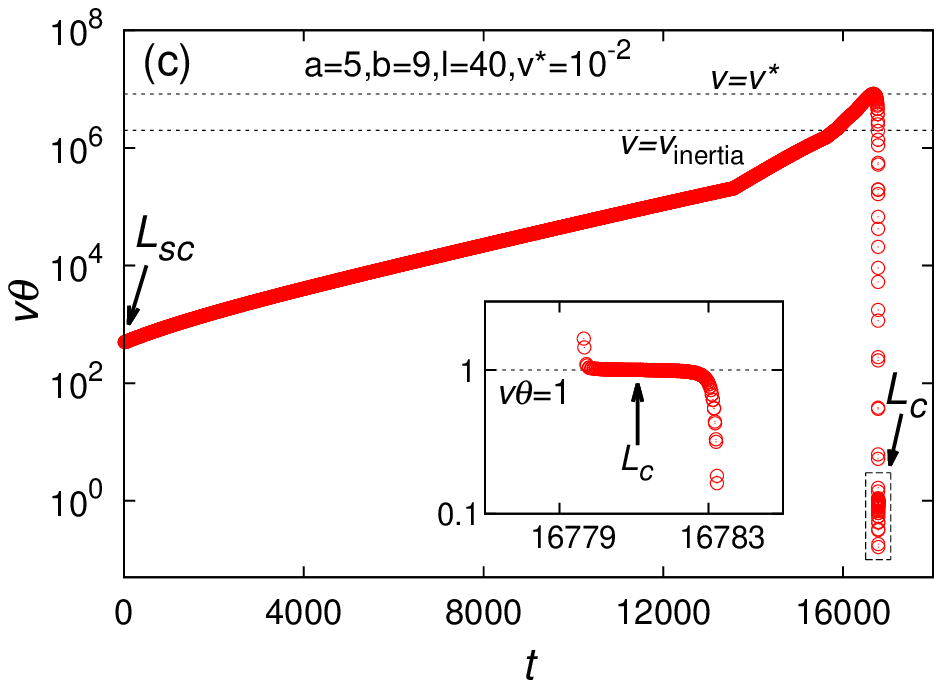}
\end{center}
\caption{
The time evolutions of (a) the sliding velocity $v$, (b) the state variable $\theta$, and (c) the multiple of the two $v\theta$, of an epicenter block in the acceleration phase in the weak frictional instability regime. The model parameters are $a=5$, $b=9$, $c=1000$, $l=40$, $v^*=10^{-2}$ and $\nu=10^{-8}$, corresponding to the near-continuum regime. The time origin $t=0$ is taken at the point of $L=L_{sc}$. The arrows indicate the points of $L=L_{sc}=43.42$ and of $L=L_c=157$. 
%The dotted horizontal lines represent the lines corresponding to $v=v^*$ and $v=v_{inertia}$. 
The solid curves in (a) are the theoretical fitting curves, eq.(\ref{vy}) with eq.(\ref{tL3}) ($v<v^*$), and eq.(\ref{v>v*}) with eq.(\ref{tL>v*}) ($v>v^*$). % The insets are magnified views of the region around $L_{c}$.
}
\end{figure}
% The insets are magnified views of the region around $L_{c}$. 
%

 To a good precision, the maximum sliding velocity is reached when the relation $v\theta =1$ is met. Eq.(\ref{aging}) indicates that, when the condition $v\theta =1$ is met, the state variable $\theta$ takes a minimum. Meanwhile, $v\theta$ sticks to a value close to unity in this range, yielding the relation $v=1/\theta$. It means that the sliding velocity $v$ takes a maximum at the point where the condition $v\theta =1$ is met.

 In Fig.10(a), we show the time evolution of the number of simultaneously moving blocks, {\it i.e.\/}, the nucleus size $L$. The data exhibit a sharp peak at which the number of simultaneously moving blocks becomes maximum. This point was taken in subsection A as our tentative criterion of $L_c$ ($L'_c$).  At or very close to this point, the epicenter block ceases to move (the double arrow in the figure), beyond which the group of simultaneously moving blocks are detached into two parts, each part propagating in opposite directions. 

 The point where $v\theta$ takes a value unity and the epicenter-block sliding velocity reaches its maximum, might also be used as a reasonable criterion of $L_c$. This definition of $L_c$ tends to yield a $L_c$-value somewhat smaller than our previous definition of $L_c$ ($L_c^\prime$), {\it i.e.\/}, the maximum of the number of the simultaneously moving blocks. One justification of the new criterion might be the observation that the epicenter-block motion in the time range after $v\theta $ levels off around $v\theta=1$ has already become similar to the one observed in a typical block motion in the high-speed rupture phase. In this sense, the high-speed rupture has already set in in the epicenter region when the epicenter blocks satisfies the relation $v\theta \simeq 1$. Thus,  in the following, we adopt  as our criterion of $L_c$ the relation $v\theta=1$ being reached at the epicenter block. This point agrees with the point of $\theta$ taking a minimum, or $v$ taking a maximum. In fact, the $L_c$-values indicated in Figs.1,4,5,8 and 9 above were the ones defined in this way unless otherwise stated. 

 In earthquake dynamics, there generally exist two different types of velocities. One is the fault sliding velocity (particle velocity), corresponding in our model to the block sliding velocity $v$. The other is the rupture-propagation velocity (phase velocity), corresponding in our model to the propagation speed of the rim of the rupture zone $v_r$. In the nucleation process, the latter also coincides with the growth speed of the nucleus size ($\sim $ half of it). Although the definition of the rupture-propagation velocity $v_r$ is somewhat obscure in the discrete BK model especially in the strongly discrete case, it might be well-defined in the near-continuum case as a (coarse-grained) growth rate of the rim of the nucleus. Namely, if the rim of the nucleus moves from the block $j$ to $j+\Delta j$ in a unit time interval, the rupture-propagation velocity might be defined by $v_r=1/\Delta j$. We show in Fig.10(b) the time evolution of the rupture-propagation velocity $v_r$ computed in this way in the near-continuum case. 

 In the acceleration phase between $L=L_{sc}$ and $L=L_c$, we identify two characteristic points where the block motion appears to change its behavior. One is the point where the epicenter-block sliding velocity exceeds the crossover velocity $v^*$, across which the $a$-term gradually changes its character. The other is the crossover velocity $v_{inertia}$ at which the inertia effect becomes important. The inertia effect as meant here is borne by the first term of the r.h.s. of the equation of motion (\ref{eq-motion}) or (\ref{eq-motion2}). This term tends to suppress the rapid acceleration, giving rise to the saturation and the subsequent drop of the sliding velocity $v$. These two characteristic points also manifest themselves in our theoretical analysis of \S IV below. 

\begin{figure}[ht]
\begin{center}
\includegraphics[scale=0.9]{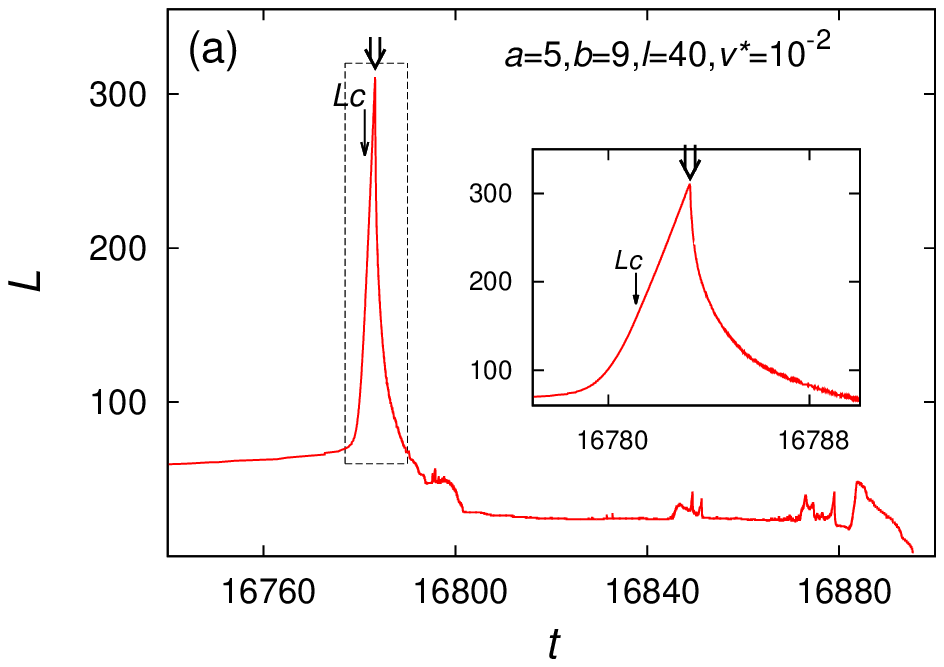}
\includegraphics[scale=0.8]{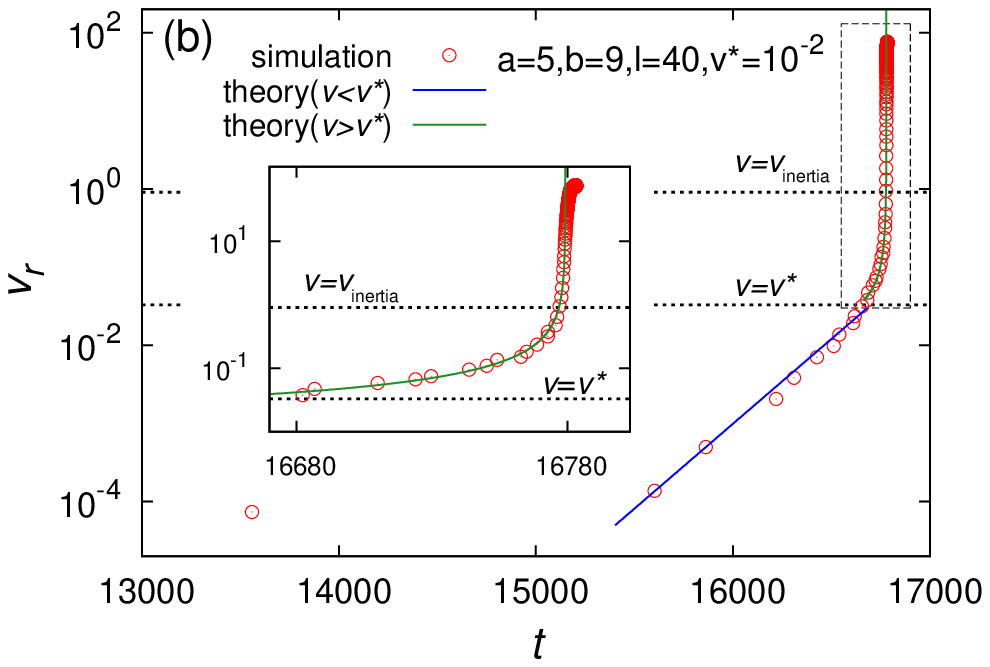}
\end{center}
\caption{
The time variations of (a) the total number of simultaneously moving blocks (the nucleus size) $L$, and of (b) the rupture propagation velocity $v_r$.  The model parameters are $a=5$, $b=9$, $c=1000$, $l=40$, $v^*=10^{-2}$ and $\nu=10^{-8}$ corresponding to the near-continuum case (the same as in Figs.7 and 9). The arrow indicates the point of $L=L_c$, while the double arrow indicates the point where an epicenter block stops. The dotted horizontal lines in (b) represents the lines $v=v^*$ and $v=v_{inertia}$. The solid curves in (b) are the theoretical fitting curves, eq.(\ref{vr}) with eq.(\ref{tL3}) ($v<v^*$) and eq.(\ref{vr>v*}) with eq.(\ref{tL>v*}) ($v>v^*$).
}
\end{figure}

One sees from Fig.10(b) that the rupture-propagation velocity $v_r$ grows exponentially with the time until around $v\simeq v^*$, beyond which it grows faster than exponential (super-exponential). By contrast, as can be seen from Fig.9(a), the epicenter-block sliding velocity exhibits a faster-than-exponential growth even in the acceleration phase at $v<v^*$. Namely, the sliding-velocity accerelation dominates over the nucleation-size expansion. Meanwhile, the super-exponential rapid growth of both the sliding velocity and the rupture-propagation velocity tends to be suppressed beyond the crossover velocity $v_{inertia}$, which is caused by the inertia effect borne by the first term of eq.(\ref{eq-motion2}).

 In Fig.11, we show (a) the epicenter-block sliding velocity $v$, and (b) the rupture propagation velocity $v_r$, versus the nuclear size $L$ normalized by $L_{sc}$, $L/L_{sc}$, instead of the time $t$. The theoretical curves to be derived in \S IV are also shown in the figure for comparison. From this, the changes of behavior at $v\simeq v^*$ and at $v\simeq v_{inertia}$ are eminent. The comparison with the analytical results are sometimes more direct in this form.

% The motion of the blocks located in the outer region of the nucleus is also shown in Fig., the ones located *, * and * blocks away from the epicenter block. These blocks begin to move with some time lag to the epicenter block, reaches their maxima of about $v\simeq 1$, and then deacceleration and stops slightly after the epicenter block stops. The maximum velocity tends to become larger for the block located in the outer region of the nucleus, but the overall behavior is quite similar. In any case, the block sliding velocity within the nucleus becomes pretty large during the acceleration phase, comparable to the sliding velocity at the main rupture. % This observation gives us some hope to detect the nucleation process during the acceleration phase. The remaining issue is the duration time of the acceleration phase (if it is too short, we have practically no time to detect it and to prepare for the coming mainshock!), and the sliding velocity reaches its maximum at which point in the acceleration phase. 

%
\begin{figure}[ht]
\begin{center}
\includegraphics[scale=0.8]{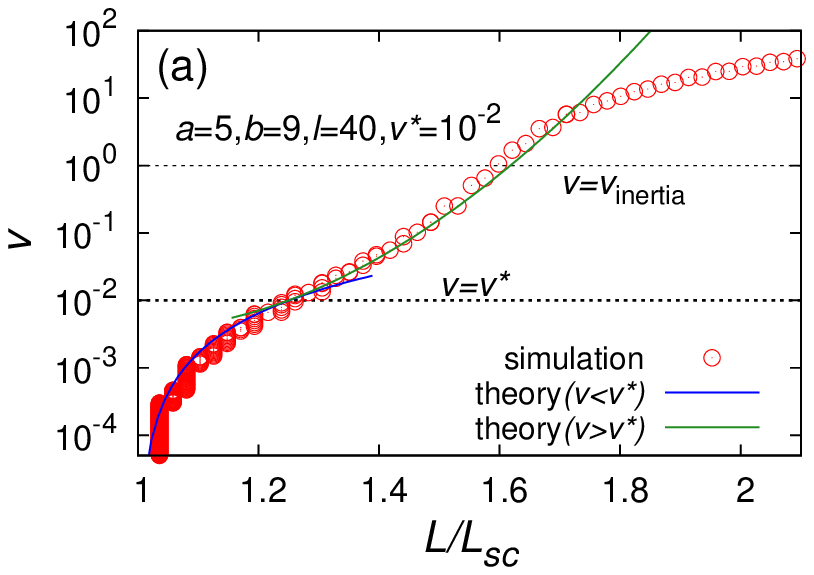}
\includegraphics[scale=0.8]{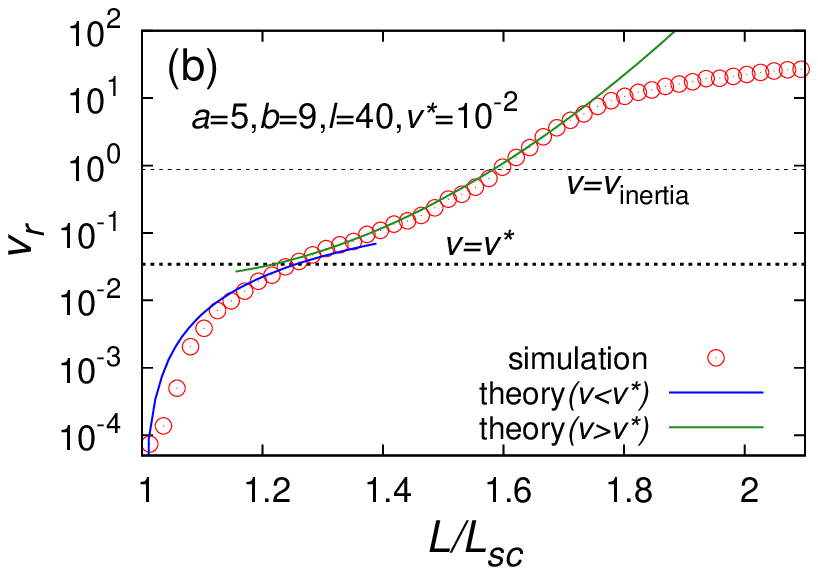}
\end{center}
\caption{
(a) The epicenter-block sliding velocity $v$, and (b) the rupture-propagation velocity $v_r$, plotted versus the nucleus size divided by the nucleation length, $L/L_{sc}$, in the acceleration phase in the weak frictional instability regime. The model parameters are $a=5$, $b=9$, $c=1000$, $l=40$, $v^*=10^{-2}$ and $\nu=10^{-8}$, corresponding to the near-continuum case. The time origin $t=0$ is taken at the point of $L=L_{sc}$. The dotted horizontal lines represent the lines corresponding to $v=v^*$ and $v=v_{inertia}$. The theoretical fitting curves are also shown, (a) eq.(\ref{vy}) ($v<v^*$) and eq.(\ref{v>v*}) ($v>v^*$), and (b) eq.(\ref{vr}) ($v<v^*$) and eq.(\ref{vr>v*}) ($v>v^*$).
}
\end{figure}

\subsection{D. The acceleration phase of the strong frictional instability regime}

Next, we study how the dynamics evolves during the acceleration phase for the case of the strong frictional instability. Remember that the model in the strong frictional regime lacks in the quasi-static initial phase.

  The block motion here turns out to be similar to that of the weak frictional instability regime with a stronger discreteness. In Fig.12, we show the time evolutions of (a) the sliding velocity $v$, (b) the state variable $\theta$, and (c) the multiple of the two $v\theta$, of an epicenter block in the acceleration phase in the strong frictional instability regime. The parameters are taken to be $a=5$, $b=40$, $c=1000$, $l=4$, $v^*=10^{-2}$ and $\nu=10^{-8}$. In the event of Fig.12, the nucleation length $L_c$ is $L_c=1$, {\it i.e.\/}, the condition $v\theta =1$ has been met during the one-block motion. Note that, in contrast to the weak-frictional instability case, this one-block motion is already irreversible.

 One noticeable feature appears at an earlier stage of the high-speed rupture phase in the strong frictional instability regime. Namely, the block velocity often exhibits prominent oscillations with the time. The maximum sliding velocity realized at each oscillation is pretty high, comparable to that of a mainshock. In the inset of Fig.12, we show an expanded view around $L_c$. Such an oscillatory behavior is rarely seen in the case of the weak frictional instability.

 A closer look of the color plot in the inset of Fig.3 might reveal that such a velocity oscillation of the block velocity is borne by the propagation and the multiple reflections of the rupture front originally ejected at $L=L_c$ from the epicenter block. This rupture front propagates along the fault with an elastic-wave velocity $\sim l$, eventually becomes a rupture front of a mainshock. In the early stage of the high-speed rupture phase, this propagating rupture front is reflected every time it reaches a neighboring block, generating the second, third, $\cdots$ rupture fronts, forming an oscillatory pattern. The period of oscillation should be given by $2/l$, which, in the example of Fig.12, yields 0.5. This period is expected to be independent of the parameters like the plate loading velocity $\nu$ or the crossover velocity $v^*$, while it might increase weakly with the frictional parameter $b$ because of the slowing-down effect due to the friction. Because of the friction, the velocity of the subsequent rupture-front propagation tends to be reduced, making the oscillation period a bit longer at a later time. We find that such expectations are consistent with the observation. For example, in an example shown in Fig.12, the observed oscillation period is 0.8-1.5, a bit longer than the expected value of 0.5. 

 Thus, in the strong frictional instability regime, the beginning of the high-speed rupture phase seems to be characterized by the multiple reflections of the propagating rupture front originally ejected from the epicenter site. After some time, the leading propagating rupture alone survives and propagates with an elastic-wave velocity $\sim l$ for the major part of the mainshock.

\begin{figure}[ht]
\begin{center}
\includegraphics[scale=0.73]{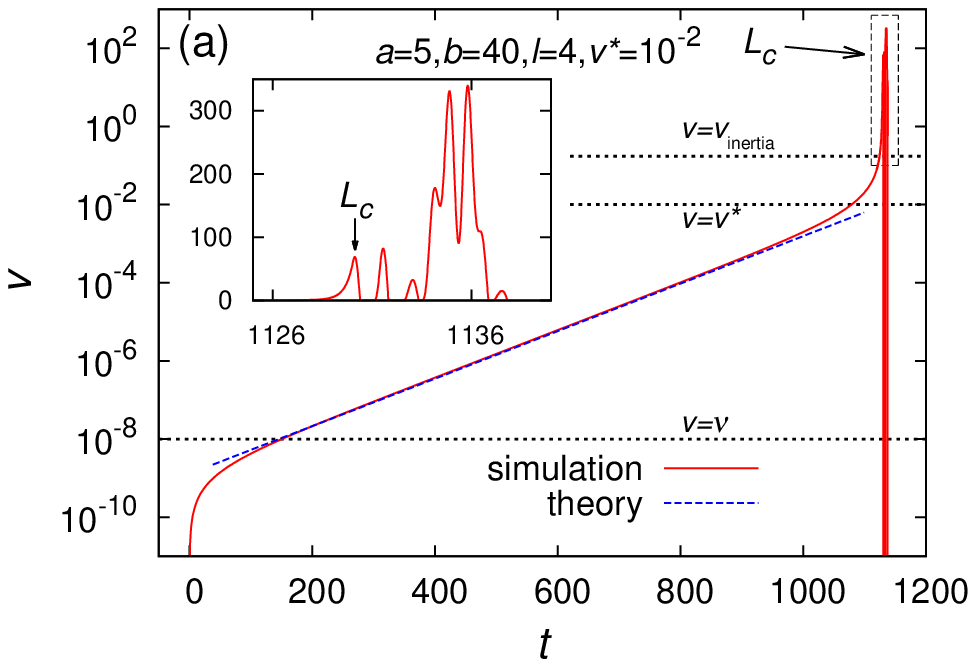}
\includegraphics[scale=0.73]{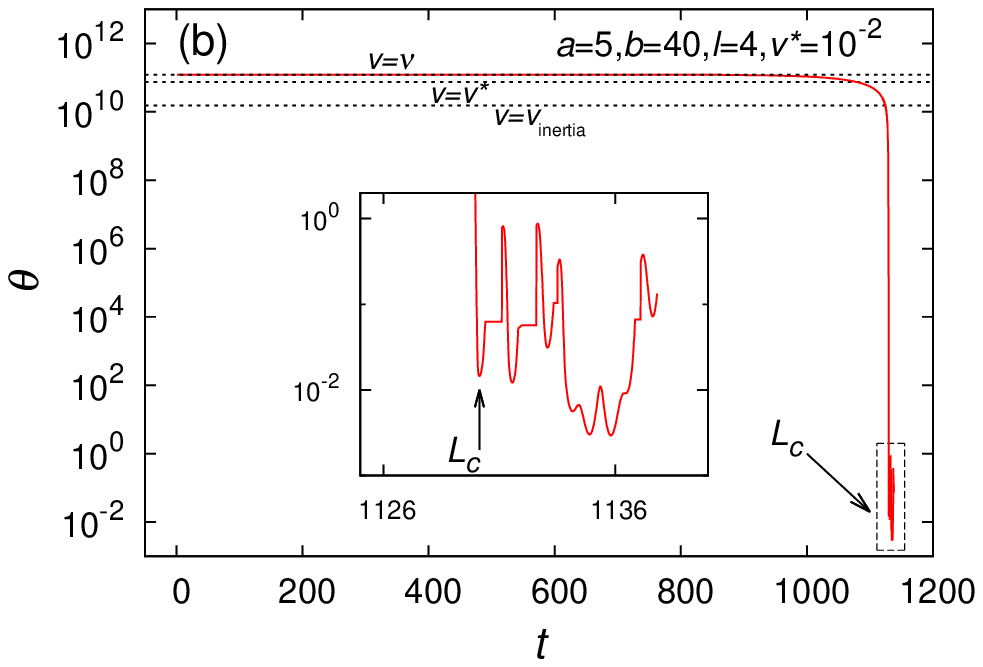}
\includegraphics[scale=0.73]{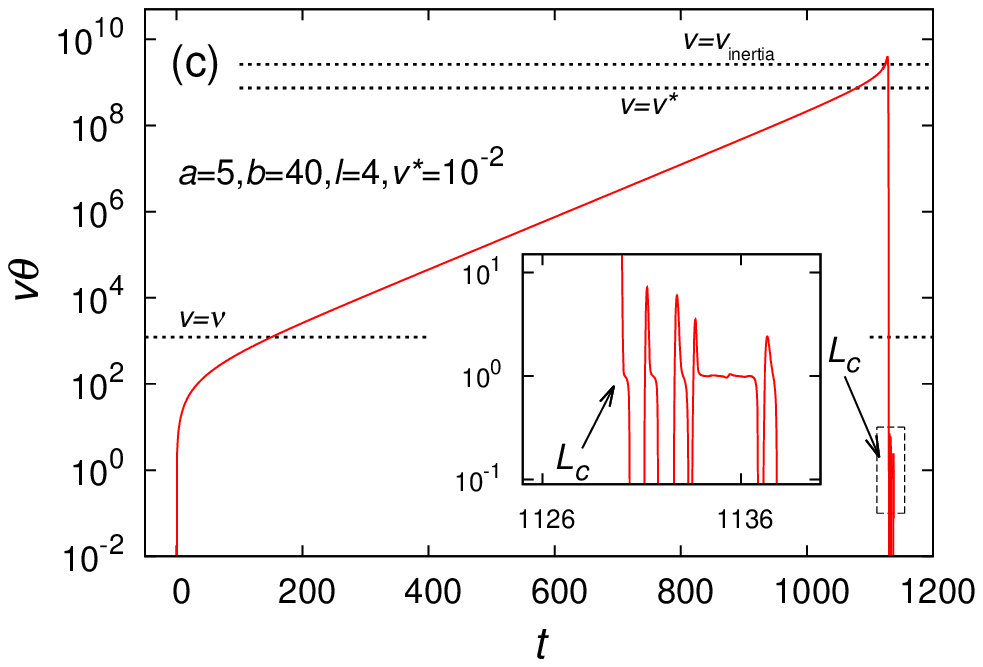}
\end{center}
\caption{
The time evolutions of (a) the sliding velocity $v$, (b) the state variable $\theta$, and (c) the multiple of the two $v\theta$, of an epicenter block in the acceleration phase in the strong frictional instability regime. The model parameters are $a=5$, $b=40$, $c=1000$, $l=4$, $v^*=10^{-2}$ and $\nu=10^{-8}$. The time origin $t=0$ is taken at the point where the epicenter begins to move. The arrow indicates the point of $L=L_c$. The dotted horizontal lines represent the lines corresponding to $v=\nu$, $v=v^*$ and $v=v_{inertia}$. The solid line in (a) is the theoretical curve for fixed $L$, eq.(\ref{vL}). The insets are magnified views of the region around $L_{c}$.
}
\end{figure}

\section{IV. Analytical treatments}

 In this section, we wish to report on the results of our analytical treatments of the dynamical properties of the model, those based on either the perturbation theory [A]-[D], or on the mechanical stability analysis [E].  Readers interested only in the simulation results might skip to \S V.

\subsection{A. Perturbation theory}

 We begin with the equation of motion (\ref{eq-motion2}) for the velocity variable $v_i$. As mentioned, the low plate pulling speed $\nu$ provides an extremely small number, say, $\nu \sim 10^{-7}-10^{-8}$. Furthermore, throughout the nucleation process, the state variable $\theta$ tends to keep a very large value of order $1/\nu$. Then, it might be convenient to introduce the reduced state variable of order unity, $\tilde \theta$, defined by $\tilde \theta \equiv \nu \theta$. One gets a set of equations of motions in terms of $v_i$ and $\tilde \theta_i$,
\begin{eqnarray}
\frac{{\rm d}^2v_i}{{\rm d}t^2} &+& \frac{a}{v^*+v_i} \frac{{\rm d}v_i}{{\rm d}t} - l^2(v_{i+1}-2v_i+v_{i-1})+ (1-b) v_i \nonumber \\ 
&=& \nu\left( 1-\frac{b}{\tilde \theta_i}\right) , \\ 
\frac{{\rm d}\tilde\theta_i}{{\rm d}t} &=& \nu - v_i\tilde \theta_i ,
\end{eqnarray}

 Then, we introduce the ``first Fourier-mode approximation'', which states that for the most part of the nucleation process the spatial form of observables, {\it i.e.\/}, the $i$-dependence of the block sliding velocity $v_i$ or the block displacement $u_i$, is given by that of the first Fourier mode. Namely, when the total $L$ blocks from $i=1$ to $i=L$ are moving, $v_i$ or $u_i$ is proportional to $\sin \left( \frac{\pi}{L+1} i \right)$.  In this approximation, it is implicitly assumed that the nucleus keeps a highly symmetrical form and the central block is an epicenter block.

 We show in Fig.13 the spatial form of the block sliding velocity $v_i$ observed  in our numerical simulations at several representative points of a typical nucleation process in the weak frictional instability regime, including (a) the initial phase, (b) the acceleration phase at $v<v_{inertia}$, and (c) the acceleration phase at $v>v_{inertia}$, together with the first Fourier-mode forms. Except for the time range beyond $v=v_{inertia}$ close to $L_c$ of Fig.(c), this approximation turns out to be reasonably good, allowing one to reproduce the motion of an arbitrary block within the nucleus by tracing only the motion of the central block $i=\frac{L+1}{2}$ (for odd $L$).

\begin{figure}[ht]
\begin{center}
\includegraphics[scale=0.73]{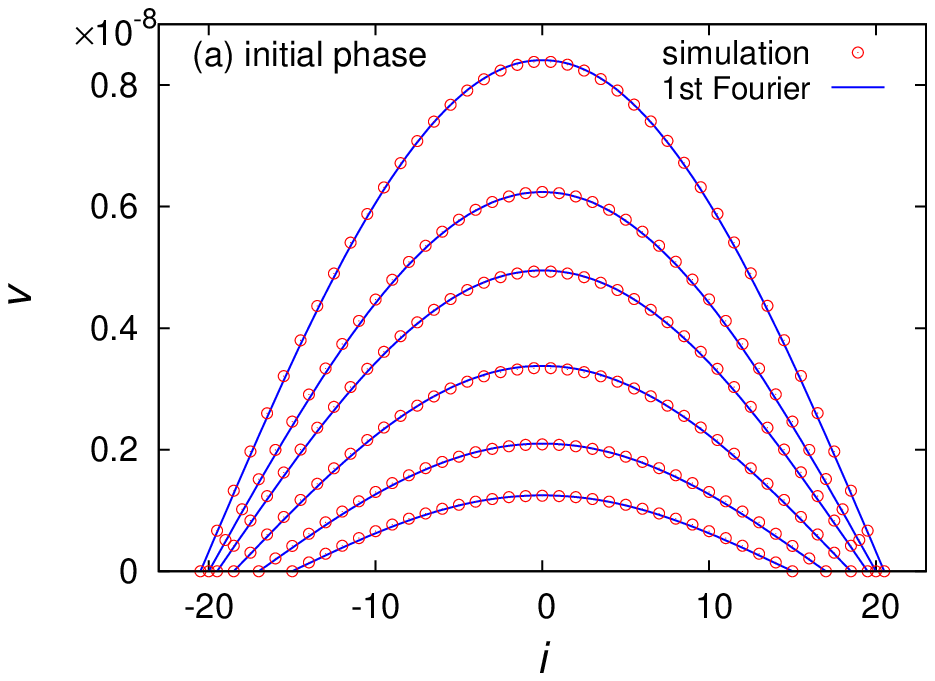}
\includegraphics[scale=0.73]{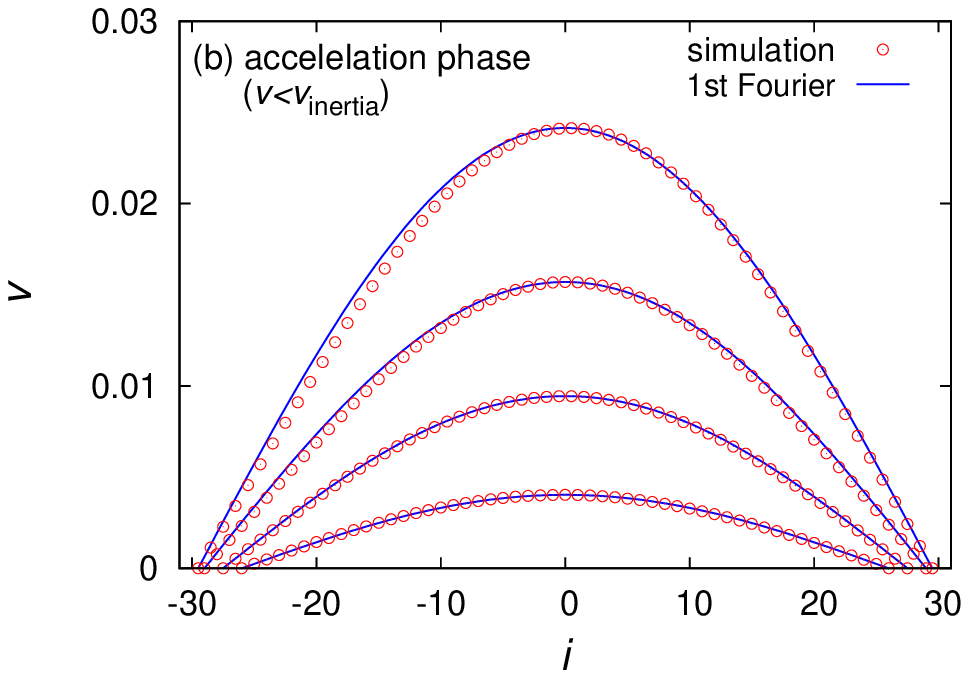}
\includegraphics[scale=0.73]{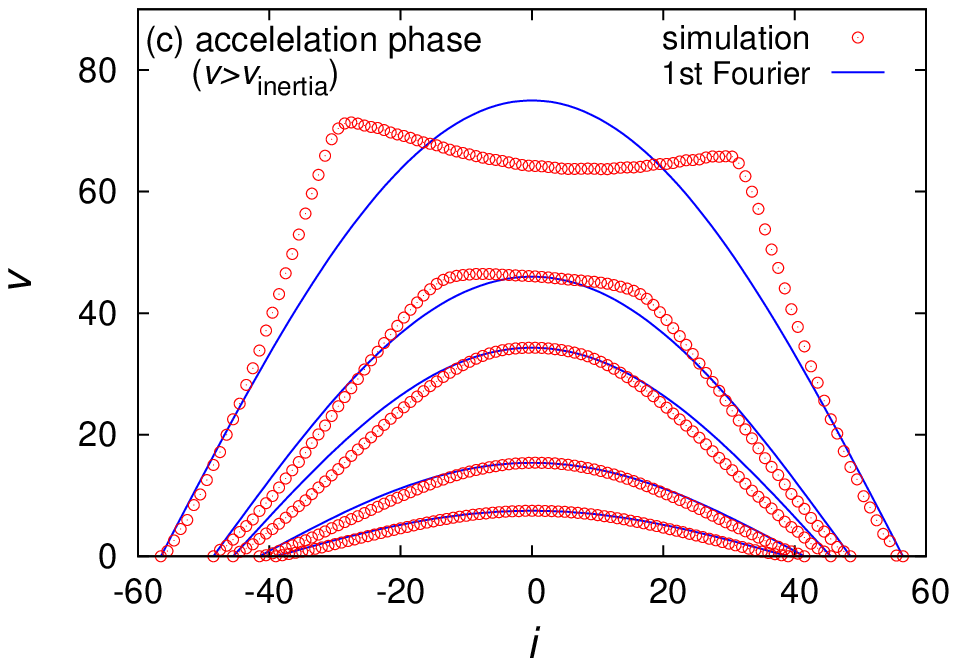}
\end{center}
\caption{
The time evolutions of the spatial pattern of the block-sliding velocity within the nucleus, (a) in the initial phase, (b) in the acceleration phase at $v<v_{inertia}$, and (c) in the acceleration phase at $v>v_{inertia}$. The model parameters are $a=5$, $b=9$, $c=1000$, $l=40$, $v^*=10^{-2}$ and $\nu=10^{-8}$. The solid curves represent the ones expected in the first Fourier-mode approximation where the height (the maximum velocity) is adjusted to the observed value.
}
\ref{nucleation}
\end{figure}

 Under this first Fourier-mode approximation, the equations of motion for the central block is given by
\begin{eqnarray}
\frac{{\rm d}^2v}{{\rm d}t^2} &+& \frac{a}{v^*+v} \frac{{\rm d}v}{{\rm d}t} + (\xi_L-b) v = \nu (1-\frac{b}{\tilde \theta}) , \\ 
\frac{{\rm d}\tilde \theta}{{\rm d}t} &=& \nu - v\tilde \theta ,
\end{eqnarray}
%s
where $\xi_L$ has been given in eq.(\ref{xiL}), and the subscript $i$ is dropped here and below. 

 Now, we expand $v$ and $\tilde \theta$ with respect to a small quantity $\nu$, {\it i.e.\/}, we perform a perturbation expansion in $\nu$,
\begin{eqnarray}
v = v^{(0)} + \nu v^{(1)} + \cdots, \\ 
\tilde \theta = \tilde \theta^{(0)} + \nu \tilde \theta^{(1)} + \cdots.
\end{eqnarray}
At the zeroth order in $\nu$, one gets a set of equations
\begin{eqnarray}
\frac{{\rm d}^2v^{(0)}}{{\rm d}t^2} &+& \frac{a}{v^*+v^{(0)}} \frac{{\rm d}v^{(0)}}{{\rm d}t} + (\xi_L-b) v^{(0)} = 0 , 
\label{eq-motion0} \\ 
\frac{{\rm d}\tilde \theta^{(0)}}{{\rm d}t} &=& - v^{(0)}\tilde \theta^{(0)} ,
\label{aging0}
\end{eqnarray}

 The equation (\ref{eq-motion0}) for the zeroth-order velocity $v^{(0)}$ has two types of solutions, {\it i.e.\/}, [A] $v^{(0)}=0$, and [B] $v^{(0)}\neq 0$. The solution [A] describes the situation where the block is at rest when the plate drive is tuned off ($\nu\rightarrow 0$). By contrast, the solution [B] describes the situation where the block is moving even when the plate drive is turned off. Hence, one expects that the solution [A] describes the initial phase, while the solution [B] describes the acceleration phase. In the following subsections, we analyze each case separately in some more detail.

\subsection{B. The initial phase}

 Here the block sliding velocity at the zeroth order is zero, $v^{(0)}=0$. As was seen in \S III, an eminent feature of the block motion in the initial phase is that there exist two time scales: One is a slow motion of order the loading velocity $\nu<<1$, and the other is a faster one of order unity. One way to deal with such two different time scales within the perturbative scheme might be to introduce two kinds of time variables, $\tau=\nu t$ associated with a slow motion and $t$ associated with a fast motion, and regard various observables as a function of both $\tau$ and $t$ like $v^{(1)}(\tau, t)$, $\theta^{(0)}(\tau, t)$ and $\theta^{(1)}(\tau, t)$. The original time derivative in the equation of motion is replaced by $\frac{{\rm d}}{{\rm d} t} \rightarrow \frac{\partial}{\partial t} + \nu \frac{\partial}{\partial \tau}$. Since $v^{(0)}=0$, the zeroth-order equation (\ref{aging0}) is reduced to $\frac{\partial \tilde \theta^{(0)}}{\partial t} = 0$. This means that $\tilde \theta^{(0)}$ depends only on $\tau$, not on $t$, {\it i.e.\/}, $\theta^{(0)}(t, \tau)=\theta^{(0)}(\tau)$.

 The equations of motion at $O(\nu)$ then reads as
\begin{eqnarray}
\frac{\partial^2v^{(1)}}{\partial t^2} &+& \frac{a}{v^*} \frac{\partial v^{(1)}}{\partial t} + (\xi_L-b) v^{(1)} = 1-\frac{b}{\tilde \theta^{(0)}} ,
\label{eq-motion2-1st} \\ 
\frac{\partial \tilde \theta^{(1)}}{\partial t} &+& \frac{\partial \tilde\theta^{(0)}}{\partial \tau} = 1 -  v^{(1)} \tilde \theta^{(0)} .
\label{aging-1st}
\end{eqnarray}
Since the zeroth-order quantity $\tilde \theta^{(0)}$ is bounded in the $t\rightarrow \infty$ limit, the corresponding first-order quantity $\tilde \theta^{(1)}$ needs to remain finite in the $t\rightarrow \infty$ limit in order that the perturbation analysis remains meaningful, {\it i.e.\/}, the relation $\frac{\partial \tilde \theta^{(1)}}{\partial t} =0$ is required in the $t\rightarrow \infty$ limit. This relation is met if the equality
\begin{equation}
\frac{\partial \tilde \theta^{(0)}}{\partial \tau} = 1 - v^{(1)}(t\rightarrow \infty, \tau) \tilde \theta^{(0)}
\end{equation}
holds. From eq.(\ref{eq-motion2-1st}), $v^{(1)}(t\rightarrow \infty, \tau)$ is obtained as 
\begin{equation}
v^{(1)}(t\rightarrow \infty, \tau) = \frac{1}{\xi_L-b}\left( 1 - \frac{b}{\tilde \theta^{(0)}(\tau)}\right).
\end{equation}
Substituting this into eq.(\ref{aging-1st}) and taking the $t\rightarrow \infty$ limit, one gets an equation to determine the hitherto undetermined $\tau$-dependence of $\theta ^{(0)}(\tau)$ as
\begin{equation}
\frac{\partial \tilde \theta^{(0)}}{\partial \tau} = -\frac{1}{\xi_L-b}\left( \tilde \theta^{(0)} - \xi_L \right) .
\end{equation}
This can be solved to yield,
\begin{equation}
\tilde \theta^{(0)}(\tau) = \xi_L + (\tilde \theta^{(0)}_0 - \xi_L) e^{-\frac{\tau}{\xi_L-b}} .
\end{equation}
where $\tilde \theta^{(0)}_0 = \tilde \theta^{(0)}(\tau=0)$. Substituting this into eq.(\ref{eq-motion2-1st}), one can get a full solution of eq.(\ref{eq-motion2-1st}) as
\begin{eqnarray}
v^{(1)}(&t&, \tau) = C_+e^{\lambda_+t} +  C_-e^{\lambda_-t} \nonumber \\ 
&+& \frac{1}{\xi_L-b}\left(1-\frac{b}{\xi_L+(\tilde \theta_0-\xi_L)e^{-\frac{\tau}{\xi_L-b}}} \right) , 
\label{v1}
\end{eqnarray}
where $C_\pm$ are numerical constants to be determined via the initial condition, and $\lambda_\pm$ is given by eq.(\ref{lambda}).

 While $\lambda_-$ is always negative, $\lambda_+$ is either negative or positive depending on whether $b<\xi_L$ or $b>\xi_L$. When $b<\xi_L$, both $C_\pm$ terms vanish quickly in eq.(\ref{v1}), whereas, when $b>\xi_L$, the $C_+$ term grows quickly leading to the instability. In fact, the borderline case $b=\xi_L$ represents the nucleation length $L_{sc}$ discriminating the stable nucleation process corresponding to the initial phase and the unstable nucleation process corresponding to the acceleration phase. 

 Now, from the condition $b=\xi_L=2l^2(1-\cos\frac{\pi}{L+1})+1$, we reach an analytical expression of $L_{sc}$ as
\begin{equation}
L_{sc} =\frac{\pi}{\arccos \left( 1-\frac{b-1}{2l^2} \right)} - 1 .
\label{Lsc}
\end{equation}
In the discrete BK model, the initial phase realized at $L<L_{sc}$ is ever possible only when $L_{sc}$ is greater than the lattice spacing or the block size, {\it i.e.\/}, $L_{sc} > 1$. Then, the condition of this nucleation length being greater than the block size $L_{sc} > 1$ yields the condition of the weak frictional instability, 
\begin{equation}
b<b_c=2l^2+1 ,
\end{equation}
where the quasi-static nucleation process is realizable in the discrete BK model. In other words, when $b>b_c$, $L_{sc}$ is less than the block spacing and the quasi-static nucleation process cannot be realized in the BK model due to its intrinsic discreteness. This is exactly the point discussed by Rice \cite{Rice}.  Hence, either the weak or the strong frictional instability is determined by the relation between the two parameters $b$ and $l$ only, a strong instability for $b>b_c=2l^2 +1$ and and a weak instability for $b<b_c$.

  We emphasize that the continuum limit of the model corresponds to $l\rightarrow \infty$ so that the continuum limit of the BK model with spatially homogeneous parameters always lies in the weak frictional instability regime, which accompanies the quasi-static nucleation process. Another derivation of $L_{sc}$ and $b_c$ based on the mechanical stability analysis will be given in the following subsection E. 

 We note in passing that the analytic formula of $L_{sc}$ given by eq.(\ref{Lsc})  is in excellent agreement with the $L_{sc}$-value determined numerically by artificially sopping the external loading as explained in \S IIIA. Precisely speaking, the $L_{sc}$-value determined by artificially sopping the external loading could slightly deviate from the analytical result. Two reasons of such a deviation are identified. In one, a nonzero loading speed $\nu$ sometimes causes  an ``overshooting'' giving a bias toward the instability. In the other, the spatial pattern of the block displacement and the block sliding velocity within the nucleus sometimes deviate from the one assumed in deriving the analytic form of \S IV, {\it i.e.\/}, of the first Fourier-mode form.

\subsection{C. Acceleration phase at $v<v^*$}

 Next, we perform the perturbation analysis of the acceleration phase. Here, the block motion is no longer stable nor quasi-static, but is essentially unstable and irreversible. There is no slow process so that no need to consider $\tau$. In contrast to the initial phase, the zeroth-order velocity $v^{(0)}$ describing this regime should be nonzero (the solution [B]).

 We divide our analysis of the acceleration phase into the two time regimes from the technical reason, {\it i.e.\/}, the regime of $v<v^*$ and of $v>v^*$. In this subsection C, we deal with the regime $v < v^*$. The regime $v > v^*$ will be dealt with in the next subsection D. For $v<<v^*$, eq.(\ref{eq-motion0}) reduces to the linear differential equation of the form,
\begin{equation}
\frac{{\rm d}^2v^{(0)}}{{\rm d}t^2} + \frac{a}{v^*} \frac{{\rm d}v^{(0)}}{{\rm d}t} + (\xi_L-b)v^{(0)} = 0 ,
\end{equation}
whose solution is given by
\begin{equation}
v = C_+ e^{\lambda_+ t} + C_- e^{\lambda_- t} \approx C_+ e^{\lambda_+ t},
\label{v-acceleration}
\end{equation}
where $\lambda_\pm$ has been given by eq.(\ref{lambda}). Since $b>\xi_L$ in the acceleration phase, $\lambda_+$ is positive leading to the instability. The time evolution of the state variable $\tilde \theta^{(0)}$ is given by
\begin{eqnarray}
\tilde \theta^{(0)} &=& \tilde \theta^{(0)}_0 \exp \left[- \left( \frac{C_+}{\lambda_+}(e^{\lambda_+ t}-1)+\frac{C_-}{\lambda_-}(e^{\lambda_- t}-1) \right)\right] \nonumber \\ 
 &\approx& \tilde \theta^{(0)}_0 \exp \left[-\frac{C_+}{\lambda_+}(e^{\lambda_+ t}-1) \right] . 
\end{eqnarray}

 In the analysis, the size of the nucleus $L$ is assumed to be fixed. Of course, an important part of the nucleation process, particularly in the unstable acceleration phase, is how the nucleus size $L$ expands with the time and how various observables evolve under the spatial expansion of the nucleus. In order to deal with such a nucleus expansion, we need additional information about the condition under which the block contingent to the moving blocks located at the rim of the nucleus begins to move. This condition actually depends on the stress state of the block assembly at the beginning of the nucleation process in question, which was basically set by the previous large event preceding the event in question. 

 We find from our numerical simulations that, in the steady state of an earthquake sequence, the excess stress $\Delta F$, which is defined as the elastic-force difference at a given block between the initial value at the beginning of the nucleation process and the threshold value at which that block eventually begins to move involved into the nucleation process, is more or less constant over blocks involved in a given event, even though this quantity is scattered considerably over various events in an event sequence. This feature originates from the fact the stress distribution after a large event tends to be flat over blocks involved in this event. 

 Equivalently, the threshold displacement $\Delta u$, which is defined as the displacement that the block located at the rim of the nucleus exhibits in order for the neighboring block initially at rest begins to move, also turns out to be more or less constant over blocks. In fact, there is a relation $\Delta F=l^2\Delta u$. In Fig.14, we show typical distributions of $\Delta u$ divided by its average over blocks involved in a given event $\overline{\Delta u}$, $\Delta u/\overline{\Delta u}$, for various parameter sets.  The data for each parameter set is an average over $10^4$ events. As can be seen from the figure, $\Delta u/\overline{\Delta u}$ tends to obey a common distribution characterized by a single-peak structure, suggesting that the approximation to regard $\Delta u$ (or $\Delta F=l^2 \Delta u$) to be constant over blocks involved in an event may not be so bad. 

\begin{figure}[ht]
\begin{center}
\includegraphics[scale=0.8]{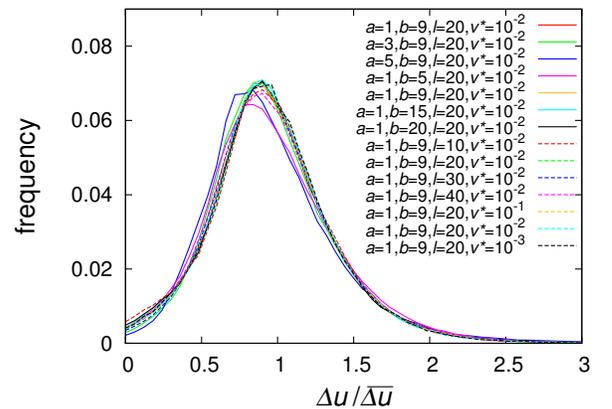}
\end{center}
\caption{
The distribution of the threshold displacement $\Delta u$ divided by its average over blocks involved in a given event $\overline{\Delta u}$, $\Delta u/\overline{\Delta u}$, for various parameter sets given in the legend. The other parameters are fixed to $c=1000$ and $\nu=10^{-8}$.
}
\end{figure}

 For convenience of the description, we introduce the reduced time variable $t^\prime $ for which the time origin $t^\prime =0$ is taken at the point where the $L$-block movement begins. In the symmetric block motion of the first Fourier-mode type we are considering here, the two blocks contingent to the nucleus begin to move entering into the nucleus motion of the size $L+2$ at the reduced time $t^\prime =t^\prime _L$. We consider the series of nuclear sizes $L_{sc},\ L_{sc}+2,\ \cdots L-2,\ L,\ L+2, \cdots$. Let the sliding velocity and the displacement of the central block at the transition from the $L$-block motion  to the $L+2$-block motion be $v_L$ and $u_L$. From eq.(\ref{v-acceleration}), one has
\begin{eqnarray}
v_L &=& v_{L-2}e^{\lambda_+(L)t^\prime _L}, 
\label{vL}
\\ 
u_L &=& \frac{v_{L-2}}{\lambda_+(L)}\left( e^{\lambda_+(L)t^\prime _L} -1 \right).
\label{uL}
\end{eqnarray}

 Within the first Fourier-mode approximation, the displacement $\Delta u$ of the block located at the rim of the nucleus means the displacement of $\Delta u/\sin (\frac{\pi}{L+1})$ of the central block. Hence, the the $\Delta u$-constant condition for the $L\rightarrow L+2$ transition can be given as the condition for the central block,
\begin{equation}
u_L = \frac{\Delta u}{\sin (\frac{\pi}{L+1})},
\end{equation}
which, together with eq.(\ref{uL}), yields the equation to determine $t^\prime _L$
\begin{equation}
e^{\lambda_+(L)t^\prime _L} = 1 + \frac{\lambda_+(L)} {v_{L-2}} \frac{\Delta u}{\sin (\frac{\pi}{L+1})} .
\label{tL}
\end{equation}
Eqs.(\ref{vL}) and (\ref{tL}) yield the recursion relation for $v_L$, 
\begin{equation}
v_L = v_{L-2} + \lambda_+(L) \frac{\Delta u}{\sin (\frac{\pi}{L+1})} ,
\end{equation}
which is solved as
\begin{equation}
v_L = \Delta u \left( \frac{\lambda_+(L)}{\sin (\frac{\pi}{L+1})} + \frac{\lambda_+(L-2)}{\sin (\frac{\pi}{L-1})} + \cdots + \frac{\lambda_+(L_{sc})}{\sin (\frac{\pi}{L_{sc}+1})} \right).
\end{equation}

 To proceed further, we consider the situation where $L$ is large enough, $L>>1$. 
% This is certainly true near the continuum limit. In the continuum limit, $L$ is taken to be large such that the dimensionless distance in the continuum $\tilde L=Ld$ ($d=\frac{D}{v_s/\omega} \rightarrow 0$ the dimensionless block size) is kept finite \cite{MoriKawamura08c}. To have a sensible continuum limit, one needs to set $d=1/l$ so that the continuum limit means $L\rightarrow \infty$ and $l\rightarrow \infty$ with $\tilde L=L/l$ kept finite. 
For $L>>1$, $\xi_L\simeq 1+(\frac{\pi l}{L})^2$, and
% = 1+(\frac{\pi}{\tilde L})^2$, and 
%
\begin{equation}
\lambda_+(L) \simeq \frac{\pi^2l^2v^*}{a}\left( \frac{1}{L_{sc}^2} - \frac{1}{L^2}\right) ,
% = \frac{\pi^2v^*}{a}\left( \frac{1}{\tilde L_{sc}^2} - \frac{1}{\tilde L^2}\right),
\end{equation}
where $L_{sc}$ is given by eq.(\ref{Lsc}). By replacing the summation by the integral, one gets
\begin{eqnarray}
v_L &=& \frac{\pi \Delta F v^*}{4a} \left(\left(\frac{L}{L_{sc}}\right)^2 - 2\ln \frac{L}{L_{sc}} -1 \right), \nonumber \\ 
 &=& \frac{\pi \Delta F v^*}{4a} (y^2-2\ln y-1) ,
\label{vy}
\end{eqnarray}
where we put $y\equiv L/L_{sc}$, 
%=\tilde L/\tilde L_{sc}$, 
and $\Delta F=l^2\Delta u$ is the excess stress defined above. This relation gives the sliding velocity of the central block as a function of the nucleus size $L$. Substituting this into eq.(\ref{tL}), one gets $t^\prime _L$ as
\begin{equation}
t^\prime _L = \frac{L_{sc}a}{\pi^2l^2v^*} \frac{y}{y^2-2\ln y-1} .
\label{tL2}
\end{equation}
This expression of $t^\prime _L$ tends to diverge in the limit $y\rightarrow 1$, {\it i.e.\/}, $L\rightarrow L_{sc}$. This is because, just at $L=L_{sc}$, the block motion is infinite slow in $t$. (Remember the relevant time scale has been of $O(\tau=\nu t)$ at $L\leq L_{sc}$.)

 In the continuum limit, $L$ is taken to be large such that the dimensionless distance in the continuum $\tilde L=Ld$ ($d=\frac{D}{v_s/\omega}$ the dimensionless block size) is kept finite \cite{MoriKawamura08c}. To have a sensible continuum limit, one needs to set $d=1/l$ so that the continuum limit means $L\rightarrow \infty$ and $l\rightarrow \infty$ with $\tilde L=L/l$ kept finite. As can be seen from eq.(\ref{tL2}), $t^\prime _L$ goes to zero in the continuum limit due to the factor $l$ in the denominator. This is simply because, in the continuum limit, the portion occupied by each fixed $L$ becomes infinitesimally small.

 The physically meaningful time in the continuum limit is a cumulative time $t_L\equiv t^\prime _{L_{sc}}+\cdots +t^\prime _L$, which is calculated as  
\begin{eqnarray}
t_L &=& \frac{\tilde L_{sc}^2 a}{2\pi^2 v^*} F(y;\epsilon), \label{tL3} \\ 
F(y;\epsilon) &=& \int^y_{1+\epsilon} \frac{y'}{y'^2-2\ln y'-1} {\rm d}y', 
\end{eqnarray}
where $y=L/L_{sc}=\tilde L/\tilde L_{sc}$ as above, and a small number $\epsilon$ takes care of removing the aforementioned divergence associated with the infinitely slow motion in $t$ around $L=L_{sc}$. This $t_L$ remains nonzero even in the continuum limit.

 The $t$-derivative of eq.(\ref{tL3}) yields another important quantity, {\it i.e.\/}, the rupture-propagation velocity $v_r\equiv \frac{1}{2}\frac{{\rm d}L}{{\rm d}t_L}$, 
\begin{eqnarray}
v_r = \frac{\pi^2 l v^*}{\tilde L_{sc}a} \frac{y^2-2\ln y-1}{y} .
\label{vr}
\end{eqnarray}

 The dimensionless rupture-propagation velocity appropriate in the continuum limit  $\tilde v_r\equiv v_rd = v_r/l$ is given by,
\begin{eqnarray}
\tilde v_r = \frac{\pi^2 v^*}{\tilde L_{sc}a} \frac{y^2-2\ln y-1}{y} ,
\label{tilde-vry}
\end{eqnarray}
where $y=\tilde L/\tilde L_{sc}$. 

If one compares this expression of $\tilde v_r$ with that of the sliding velocity $v$ of eq.(\ref{vy}), both $v$ and $\tilde v_r$ are proportional to $\frac{v^*}{a}$, meaning a larger-$a$ or a smaller-$v^*$ value tends to lead to the slower block sliding and to the slower nucleus expansion. Meanwhile, $v$ is proportional to $\Delta F$ in contrast to $v_r$, the latter being independent of $\Delta F$ (nor $\Delta u$). This means that the low stress state at the onset of the nucleation process tends to induce a high sliding velocity, but the rupture-propagation velocity is rather insensitive to the stress state.

 Comparison of the $y$-dependence of eqs.(\ref{vy}) and (\ref{vr}) suggests that the acceleration is relatively more suppressed in the rupture propagation than in the sliding velocity because of the factor $y>1$ in the denominator of eq.(\ref{vr}). In fact, for larger $y$, the rupture-propagation velocity and the nucleus size grow exponentially with the time $t$ since $\frac{{\rm d}y}{{\rm d}t}$ is proportional to $y$, 
\begin{equation}
\tilde v_r,\ \tilde L \propto \exp \left[ \frac{(b-1)v^*}{2a}t \right] , 
\label{tilde-vrt}
\end{equation}
for $y>>1$, which is consistent with our simulation data of Fig.10(b). By contrast, the sliding velocity grows faster than exponential, which has also been confirmed by our numerical simulations shown in Fig.9(a). Namely, the accerelation of the block sliding dominates over that of the nucleus expansion.

 In reality, $y=L/L_{sc}$ is not necessary much larger than unity in this regime. Even in this case, however, the r.h.s. of eq.(\ref{tilde-vry}) may be regarded as approximately being linear in $y$ with a modified proportionality coefficient, {\it i.e.\/}, the exponent in eq.(\ref{tilde-vrt}) modified from the original one to an effective one. In fact, the type of the fit we made in Figs.10(b) and 11(b) was made with the associated exponent as a fitting parameter.

\subsection{D. Acceleration phase at $v>v^*$}

 Now we wish to move on to the later part of the acceleration phase where the sliding velocity of the central block exceeds the crossover velocity $v^*$. In this situation, the equation of motion for the central block becomes nonlinear, and the treatment of the previous subsection does not apply in the same form. To proceed, we introduce an additional approximation of the ``overdamped approximation''.

 The l.h.s. of the equation of motion for $v$, eq.(\ref{eq-motion2}), consists of the three terms: the first ``inertia term'' proportional to the second time-derivative $\frac{{\rm d}^2v}{{\rm d}t^2}$, the second term proportional to the first time-derivative $\frac{{\rm d}v}{{\rm d}t}$, and the third term  proportional to the velocity itself $v$. In the low velocity region, the first term is much smaller in magnitude than the other two terms, and might safely be neglected (``overdamped approximation''). Our simulation results shown in Fig.15 indicate that the first term is indeed much smaller than the other two terms not only in the initial phase and in the acceleration phase at $v<v^*$, but also in the acceleration phase even at $v>v^*$ up to a certain point preceding $L_c$. The velocity at which the first term becomes comparable to the other two terms and the ``overdamped approximation'' fails gives an another crossover velocity, which we denote $v_{inertia}$. In the following, we take the convention to define $v_{inertia}$ by the $v$-value where the first inertia term grows to 10\% of the sencond first-derivative term.

\begin{figure}[ht]
\begin{center}
\includegraphics[scale=0.8]{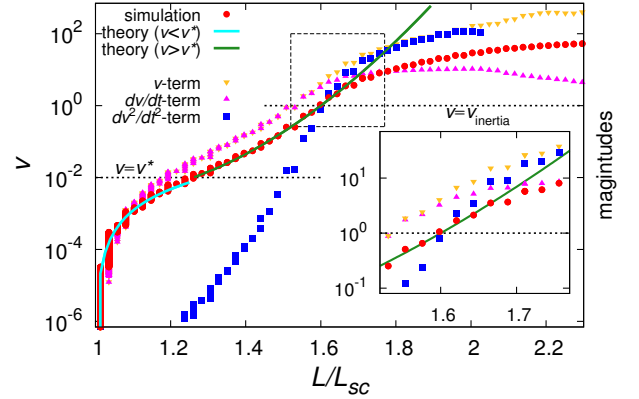}
\end{center}
\caption{
The time evolutions of the epicenter-block sliding velocity $v$ in comparison with the theoretical results  for $v>v^*$ and for $v<v^*$, and of the magnitudes of the three terms on the l.h.s. of the equation of motion, eq.(8). The first term is the ``inertia'' term containing the second time-derivative of $v_i$, the second term contains the first time-derivative of $v_i$, and the last term does not contain the time-derivative $v_i$. The horizontal lines represent $v=v^*$ and $v=v_{inertia}$. The model parameters are $a=5$, $b=9$, $c=1000$, $l=40$, $v^*=10^{-2}$ and $\nu=10^{-8}$. The inset is a magnified view. The solid curves are the theoretical fitting curves, eq.(\ref{vy}) at $v<v^*$ and eq.(\ref{v>v*}) at $v>v^*$.
}
\end{figure}

 The overdamped approximation enables one to go into the later part of the acceleration phase up to $v\simeq v_{inertia}$. Within this approximation, the sliding velocity and the displacement of the central block are calculated for a fixed $L$ to be
\begin{eqnarray}
v(t^\prime) &=& \frac{v^*}{(1+\frac{v^*}{v_{L-2}}) e^{\frac{v^*(\xi_L-b)}{a} t^\prime} - 1} , 
\label{vt} \\ 
u(t^\prime) &=& \frac{a}{\xi_L-b}\ln \left[ 1+\frac{v_{L-2}}{v^*}\left( 1-e^{\frac{v^*(b-\xi_L)}{a}t^\prime} \right) \right] .
\label{ut}
\end{eqnarray}
 Note that these expressions lead to an apparent divergence at a finite time. Of course, this is an artificial divergence caused by the ``overdamped approximation'' employed. In reality, when the velocity exceeds the crossover velocity $v_{inertia}$, the neglected ``inertia term'' becomes important suppressing the artificial divergence, and the system exhibits an entirely different behavior as can be seen from Fig.15. 

 One might describe the growth of the nucleus, {\it i.e.\/}, the time dependence of $L$, along the line of the previous subsection. Adopting the first Fourier-mode approximation and the constant-$\Delta u$ approximation, one gets from eqs.(\ref{vt}) and (\ref{ut}) the recursion relation,
\begin{eqnarray}
v_L = A_L v_{L-2} +(A_L-1)v^* ,
\end{eqnarray}
with
\begin{eqnarray}
A_L = \exp \left[ \frac{b-\xi_L}{a}\frac{\Delta u}{\sin \frac{\pi}{L+1}} \right] .
\end{eqnarray}
In the case $v_L>>v^*$ of our interest here, one may safely neglect the second term proportional to $v^* (<<v_L)$, to have 
\begin{equation}
v_L\simeq A_L v_{L-2}. 
\end{equation}
Then, one gets
\begin{eqnarray}
v_L &=& \exp [ \frac{\Delta u}{a} ( \frac{b-\xi_L}{\sin \frac{\pi}{L+1}} + \frac{b-\xi_{L-2}}{\sin \frac{\pi}{L-1}} + \cdots  \nonumber \\ 
  &+& \frac{b-\xi_{L^*+2}}{\sin \frac{\pi}{L^*+3}}) ] v^* .
\end{eqnarray}
As an initial state of the recursion relation, we take here somewhat arbitrarily the state at $v=v^*$ where the nucleus size is $L=L^*$.

In the large-$L$ limit, the summation is replaced by the integral to yield, 
\begin{eqnarray}
v = &C& v^* y^{-\frac{\pi \Delta F}{2a}} \exp \left[ \frac{\pi\Delta F}{4a} y^2\right], 
\label{v>v*}
\end{eqnarray}
with $y=L/L_{sc}=\tilde L/\tilde L_{sc}$ as above, where the constant $C$ is given by
\begin{eqnarray}
C &=& y^{* \frac{\pi \Delta F}{2a}} \exp \left[ -\frac{\pi\Delta F}{4a} y^{* 2} \right] ,
\end{eqnarray}
with $y^*=L^*/L_{sc}=\tilde L^*/\tilde L_{sc}$.
These expressions give the epicenter-block sliding velocity $v$ as a function of the nucleus size $L$ or $y$.

 The cumulative time $t_L$ is obtained as
\begin{eqnarray}
t_{\tilde L} = \frac{\tilde L_{sc}^2}{2\pi C} G(y; y^*) ,
\label{tL>v*}
\end{eqnarray}
where
\begin{eqnarray}
G(y; y^*) = \int^y_{y^*} y'^{1+\frac{\pi \Delta F}{2a}} e^{- \frac{\pi \Delta F}{4a} y'^2} {\rm d}y' .
\end{eqnarray}
The normalized rupture-propagation velocity $\tilde v_r$ is then calculated to be
\begin{eqnarray}
\tilde v_r = \frac{\pi C}{\tilde L{sc}} y^{-1-\frac{\pi \Delta F}{2a}} \exp \left[ \frac{\pi \Delta F}{4a} y^2 \right] ,
\label{vr>v*} 
\end{eqnarray}
If one compares the expression of $\tilde v_r$ with that of the sliding velocity $v$, the acceleration is relatively more suppressed in the rupture propagation than in the sliding velocity as in the case of $v<v^*$.

 Beyond $v=v_{inertia}$, the inertia effect becomes important and the system gets into the final stage of the acceleration phase, eventually approaching $L_c$. In this final time regime, the overdamped approximation fails and the equation becomes highly nonlinear so that we have no efficient analytical solution, unfortunately.

 Our numerical solution has revealed that, in this final time regime, the inertia term suppresses the acceleration, $v\theta $ drops further mitigating the acceleration, and eventually reaches the point $v\theta=1$ yielding $L_c$, which signals the onset of the high-speed rupture of a mainshock. Beyond the point $L=L_c$, the epicenter block rapidly decelerates and soon comes to a complete stop. Meanwhile, neighboring blocks begin a high-speed motion, and the system gets into the high-speed rupture phase where the rupture front propagates with the elastic wave velocity $\sim l$ in both directions.

\subsection{E. Mechanical stability analysis}

 In this subsection, we re-derive the expression of $L_{sc}$, eq.(\ref{Lsc}), based on the mechanical stability analysis, {\it i.e.\/}, from the condition of the balance between the elastic force and the friction force acting on a block \cite{Dieterich92,Scholzbook}. As mentioned, one may regard $L_{sc}$ as the length separating the stable and the unstable ruptures. When the nucleus size $L$ is less than $L_{sc}$, the rupture process is stable and reversible, whereas, when $L$ exceeds $L_{sc}$, it becomes unstable and irreversible. 

  An appropriate physical condition describing the stable/unstable sliding across $L_{sc}$ might be whether the elastic stiffness $K$, as defined by $K=\delta f_{elastic}/\delta u$ which represents a change of the elastic force $f_{elastic}$ due to an infinitesimal slip $\delta u$ of the block, is greater/smaller than the frictional weakening rate, as defined by $\delta \phi/\delta u$ which represents a change of the friction force $\phi$ due to an infinitesimal slip of the block.  If the frictional weakening rate $|\frac{{\rm d}\phi}{{\rm d}u}|$ is greater than the elastic stiffness $K$, an infinitesimal sliding $\delta u$ induces a dominance of the friction-force drop over the elastic-force drop causing a dynamical instability, {\it i.e.\/}, a slip weakening. By contrast, if the frictional weakening rate is smaller than the elastic stiffness, a further sliding is suppressed by the frictional force leading to a stable slip,  {\it i.e.\/}, a slip strengthening.

 Consider a hypothetical instantaneous process from the states ($u_i$, $v_i=0$, $\theta_i$) to ($u_i+\delta u_i$, $v_i=0$, $\theta_i+\delta \theta_i$). The aging law (\ref{aging}) entails the relation $\delta \theta_i=\delta t -\theta \delta u_i\simeq -\theta \delta u_i$. Then, the frictional-weakening rate is obtained as  
\begin{equation}
\frac{{\rm d}\phi_i}{{\rm d}u_i}=\frac{b}{\theta_i}\frac{{\rm d}\theta_i}{{\rm d}u_i}=-b. 
\end{equation}

Meanwhile, the stiffness of the $L$-block system may be given by the smallest nonzero eigenvalue of the $L\times L$ matrix $K$ defined via the relation $(\delta f_{elastic,1}, \cdots, \delta f_{elastic,L})=K (\delta u_{1}, \cdots, \delta u_{L})$ as 
\begin{equation}
K_{min}=2l^2\left( 1-\cos \frac{\pi}{L+1}\right) + 1.
\label{stiffness}
\end{equation}
The eigenfunction associated with the smallest eigenvalue $K_{min}$ just corresponds to the first Fourier mode which we employed in our approximate solution of the equation of motion. As the size of nucleus $L$ is increased, the stiffness $K_{min}$ given by eq.(\ref{stiffness}) decreases. Note that, however, even in the $L\rightarrow \infty$ limit $K_{min}$ does not vanish altogether, retaining a nonzero value, unity, in contrast to the elastic-continuum case \cite{Dieterich92,RubinAmpuero,AmpueroRubin} where $K$ vanishes as $1/L$.

 Matching $K$ and $|\frac{{\rm d}\phi}{{\rm d}u}|$, the condition of the frictional instability is obtained as
\begin{equation}
L>L_{sc} = \frac{\pi}{\arccos \left( 1-\frac{b-1}{2l^2}\right)} - 1 .
\label{L>Lsc}
\end{equation}
yielding the expression of $L_{sc}$ given by eq.(\ref{Lsc}). 

 In Fig.16, the stiffness $K$ of an epicenter block computed in the course of the nucleation process of our simulation is plotted versus the number of moving blocks $L$, together with the theoretical curve (\ref{stiffness}). The two agree very well. At an earlier stage of the slip, an inequality $K>|\frac{{\rm d}\phi}{{\rm d}u}|$ holds indicating a stable slip, while, at a certain point, an equality  $K=|\frac{{\rm d}\phi_b}{{\rm d}u}|$ is reached signaling $L_{sc}$, beyond which an opposite inequality $K < |\frac{{\rm d}\phi_b}{{\rm d}u}|$ holds indicating an unstable slip. The system then gets into the unstable acceleration phase.

\begin{figure}[ht]
\begin{center}
\includegraphics[scale=0.8]{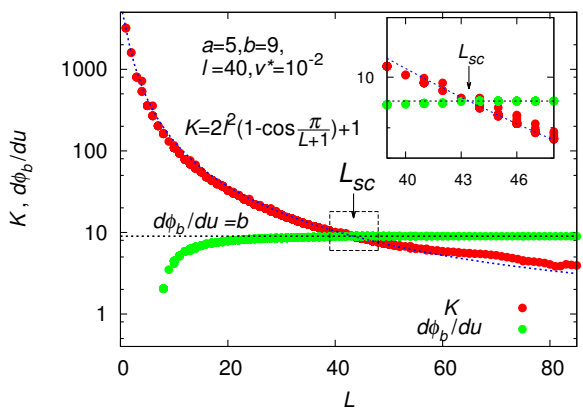}
\end{center}
\caption{
The stiffness $K$ of an epicenter block is plotted versus the number of moving blocks $L$ for a typical nucleation process. The model parameters are $a=5$, $b=9$, $c=1000$, $l=40$, $v^*=10^{-2}$ and $\nu=10^{-8}$, with $L_{sc}=43.4$. The theoretical curves of the stiffness $K$ and of the frictional-weakening rate $|\frac{{\rm d}\phi_b}{{\rm d}u}|$ are also shown: See the text for details. The nucleation length $L_{sc}$ corresponds to the crossing point of the two curves.
}
\end{figure}

 Eq.(\ref{stiffness}) might suggest that, if $b<1$, $b<K_{min}$ for any value of $L$. It means that the earthquake-like frictional instability is no longer possible in the region of $b<1$ of the model. Indeed, we observe in our simulations that, in the region of $b<1$, the model exhibits a creep-like continuous movement without showing an earthquake-like instability any more.

% What occurs near the border of the two regimes could be complex. We find that events in this regime, $b\gtrsim 1$ with $b-1<<1$, are quite unusual. In the standard earthquake-like event, a nucleation process evolves into a mainshock where a seismic rupture propagates from the nucleus with an elastic-wave velocity. In this border regime, events are almost always all-block events where all blocks rupture. This all-block event is quite unlike the standard earthquake-like event in that the rupture does not propagate with the elastic-wave velocity, but rather, it occurs almost instantaneously driven by the plate motion itself. In other words, an entire fault ruptures almost simultaneously, quite unlike the standard seismic event  realized off the border $b=1$. Such a behavior is so different from the behavior of the standard earthquake instability, and we shall not enter into this border regime any further.

\section{V Simulation results II}

 When the nucleation process precedes a mainshock, one might naturally ask how the properties of the nucleation process is related or unrelated to the properties of the ensuing mainshock itself. This question would be of particular interest in its possible connection to an earthquake forecast. In this section, we investigate the statistical properties associated with the nucleation process, {\it e.g.\/}, the nucleation lengths $L_{sc}$ and $L_c$, and the duration times of each phase of the nucleation process, averaged over many events in connection with the mainshock properties. 

 Of course, difficulties accompany such a forecast. The fault sliding is generally very slow for most part of the nucleation process, which makes the real-time detection of the nucleation process difficult. Especially in the initial phase, the fault motion is extremely slow, being of ``atomic scale'' of $\simeq 1$ [nm/s]. In the acceleration phase, the sliding velocity increases by several orders of magnitude towards the nucleation length $L_{c}$, eventually becoming comparable to the maximum sliding velocity at the main rupture. An important point here is how much time is left before the onset of the mainshock. We study in this section how the dynamics evolves during the acceleration phase in some detail, mainly for the case of the weak frictional instability relevant to the continuum limit.

\subsection{A. The nucleation lengths $L_{sc}$ and  $L_{c}$}

 As was revealed in the previous sections, the nucleation length $L_{sc}$ is determined only by the material parameters as given in eq.(\ref{Lsc}), meaning that  $L_{sc}$ cannot be used as an indicator of the size of the ensuing mainshock which may be small or large.

 What about the nucleation length $L_c$ ? Does it correlate with the final mainshock-rupture size ? We plot in Fig.17(a) the mean-$L_c$  computed in our simulations normalized by the corresponding $L_{sc}$, $L_c/L_{sc}$, versus the final rupture-zone size $L_r$ for various choices of the model parameters in the weak frictional instability regime. The $b$-value is fixed to $b=9$ while the parameters $l$, $a$ and $v^*$ are varied. The data for each parameter set is an average over $10^4$ events in the strong frictional instability regime, and  $10^5$ events in the weak frictional instability regime, except for the case of $l=10$ where the corresponding numbers are 3500 and 24000, respectively. As can be seen from Fig.17(a), the data approximately collapse onto a common curve. Since $L_{sc}$ given by eq.(\ref{Lsc}) does not depend on $a$ and $v^*$, this indicates that $L_c$ is also insensitive to $a$ and $v^*$, while its $l$-dependence is the same as that of $L_{sc}$. One also sees that $L_c$ tends to be independent of $L_r$ except for smaller events, implying that one cannot predict the size of the upcoming mainshock even from the information of $L_c$.

 We examine the $b$-dependence of $L_c/L_{sc}\equiv r$,  and plot in Fig.17(b) the mean $L_c/L_{sc}$-value versus $b$ for various $l$-values, including not only the weak frictional instability regime but also the strong frictional instability regime. As can be seen from Fig.17(b), $L_c/L_{sc}$ exhibits a nontrivial $b$-dependence accompanied by a cusp-like change of behavior at $b=b_c$ discriminating the weak and the strong instability regimes. The data in the weak frictional instability regime tend to increase almost linearly with $b$, lying on a common line even for different $l$, while those in the strong frictional instability regime tend to decrease with $b$. We find that the data in the weak frictional instability regime of $b<b_c$ exhibits a near-linear $b$-dependence well fittable by the relation $r(b) = L_c/L_{sc} \simeq 0.1b+4.4$.

\begin{figure}[ht]
\begin{center}
\includegraphics[scale=0.75]{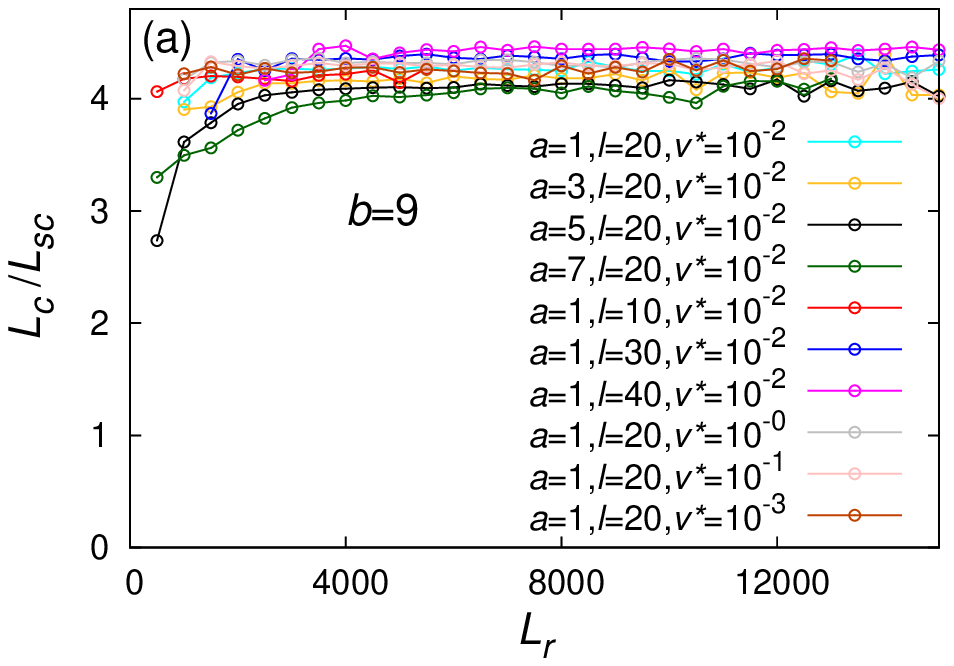}
\includegraphics[scale=0.75]{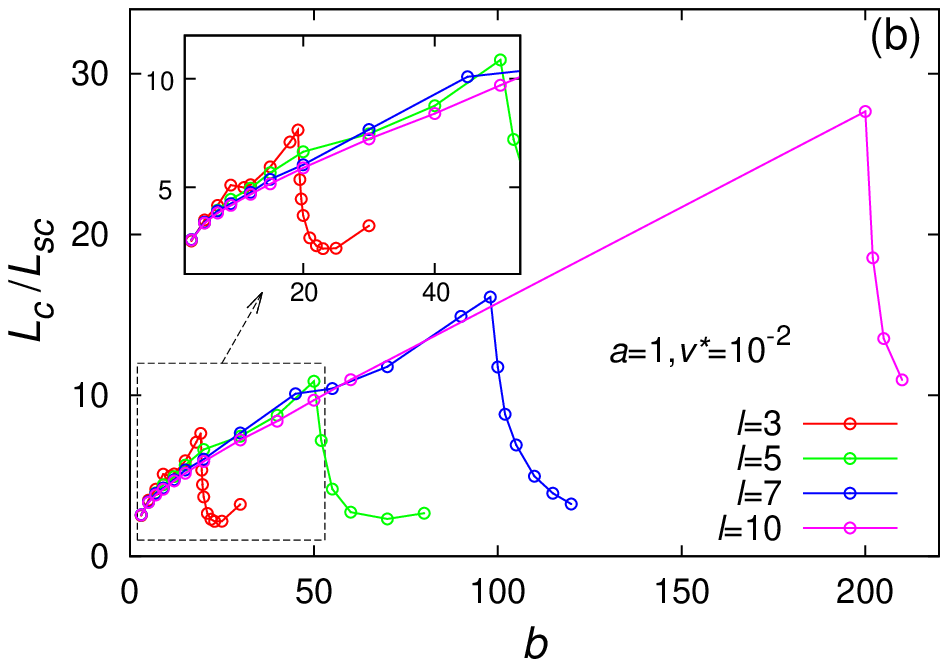}
\end{center}
\caption{
(a) The mean nucleation length $L_c$ divided by $L_{sc}$, $L_c/L_{sc}$, plotted versus the rupture-zone size $L_r$ for various parameter sets ($a$, $l$, $v^*$) in the weak frictional instability regime. The other parameters are $b=9$, $c=1000$ and $\nu=10^{-8}$. (b) The mean $L_c/L_{sc}$ plotted versus the friction parameter $b$ for several value of $l$. The other parameters are set to $a=1$, $c=1000$, $v^*=10^{-2}$ and $\nu=10^{-8}$. The inset is a magnified view of the small-$b$ region.
}
\end{figure}

\subsection{B. The duration times of each nucleation phase}

 Next, we consider the duration times of each stage of the nucleation process, including that of the initial phase $T_\alpha$ ($L< L_{sc}$), of the acceleration phase $T_\beta$ ($L_{sc}<L<L_c$), and of the high-speed rupture phase $T_\gamma$ ($L>L_c$). The ultimate utility of the nucleation phenomenon may be forecasting the upcoming mainshock. As mentioned, practical detection, if any, would become possible only in the acceleration phase. Since the system has already been beyond the ``no-return'' point, a mainshock should already be ``deterministic'' there. The remaining problem is how much time is left.

 We tentatively set the detectable sliding velocity of the nucleus motion $v =10^{-4}=10^4\nu$ which corresponds in real unit to $\simeq 10^{-2}$ [mm/sec]. Then, the time interval between the point of $v=10^{-4}$ and the point of $L=L_c$ (the onset of a mainshock) is denoted by $T_\beta^\prime$. This $T_\beta^\prime$ would give a realistic measure of the remaining time available for a mainshock forecast.

 In Fig.18(a), we show the duration times ($T_\alpha$, $T_\beta$, $T_\beta^\prime$ and $T_\gamma$) for the case of the weak frictional instability versus the associated final rupture-zone size $L_r$. The averaged number of events are the same as those of Fig.17, except for the case of $v^*=10^{-4}$ where the corresponding number is 225. Quite naturally, the duration time of the mainshock itself, $T_\gamma$, gets longer for a larger mainshock. By contrast, the duration times of the nucleation process $T_\alpha$, $T_\beta$ and $T_\beta^\prime$ are nearly independent of the size of the ensuing mainshock. This observation means that it is again hard to predict the size of the ensuing mainshock based on the duration times of the nucleation process. A closer look of the data reveals that there is even a weak anti-correlation between the duration time of the initial phase $T_\alpha$ and the size of the ensuing mainschok. Namely, $T_\alpha$ tends to be a bit shorter for larger earthquakes, though the tendency is not pronounced.

 In Fig.18(b), we plot the mean duration times averaged over all $L_r$ versus $b$ in the main panel, and versus $a$ in the inset. One sees from the figure that the duration times depend on $b$ and $a$ only weakly. In Fig.18(c), we plot these mean duration times versus $v^*$ in the main panel, and versus $1/l$ in the inset. One sees from the main panel that the duration times $T_\alpha$ and $T_\gamma$ depend on $v^*$ only weakly, but the duration times $T_\beta$ and $T_\beta ^\prime$ depend on $v^*$ rather sensitively, increasing with decreasing $v^*$.  For $v^*=10^{-4}$, $T_\beta$ is greater than $T_\gamma$ by factor of 700, while $T_\beta ^\prime$ by factor of 20.  For smaller $v^*$, $T_\beta^{\prime}$ could be even longer, although the saturating behavior seems to set in for $v^*\lesssim 10^{-4}$. Unfortunately,  taking the data for $v^*\leq 10^{-5}$ is beyond our present computational capability.  The $1/l$-dependence of these duration times shown in the inset turns out to be rather weak. We then conclude that the remaining time available for a mainshock forecast could be longer than the mainshock duration time by one or two orders of magnitude, but perhaps not much longer than that.

\begin{figure}[ht]

\begin{center}
\includegraphics[scale=0.71]{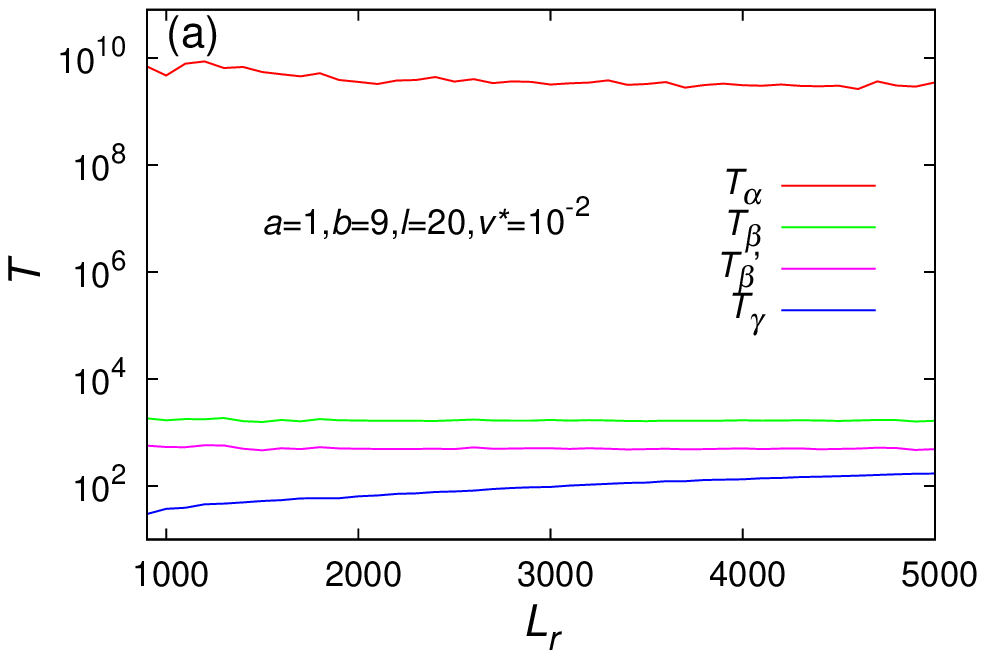}
\includegraphics[scale=0.71]{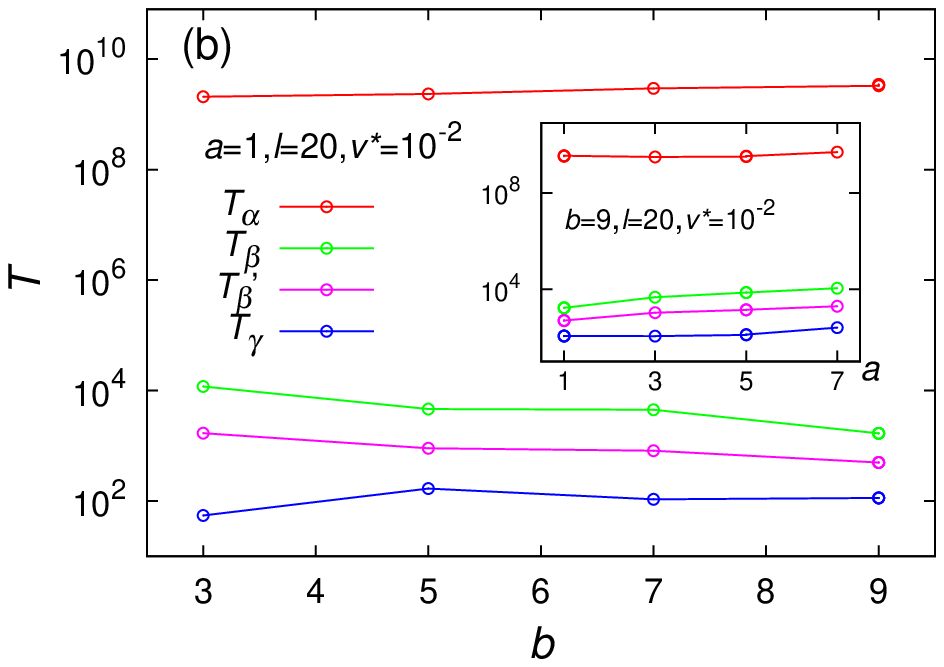}
\includegraphics[scale=0.71]{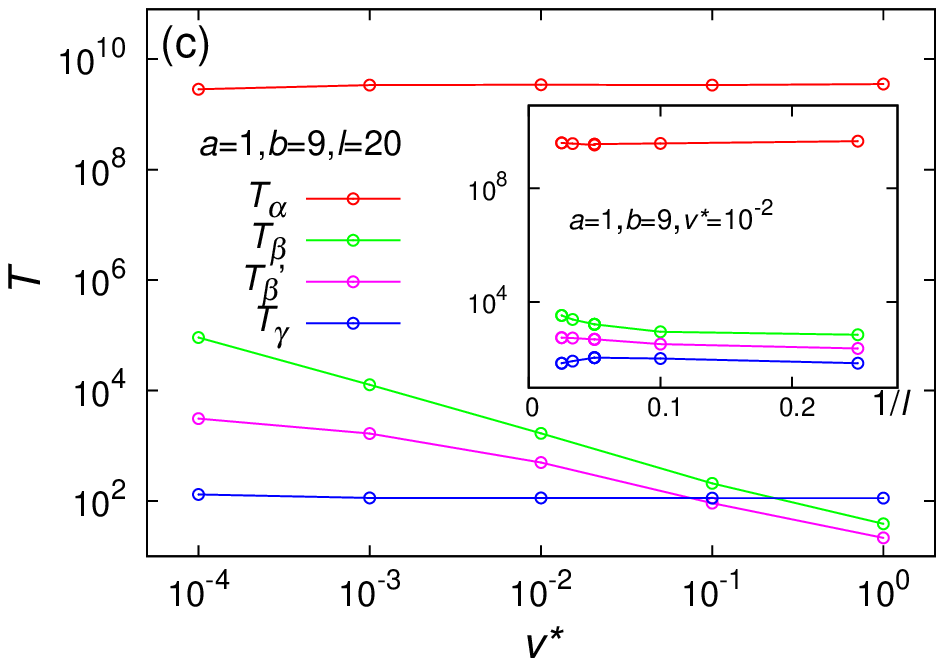}
\end{center}
\caption{
(a) The duration times, $T_\alpha$, $T_\beta$, $T_\beta^\prime$ and $T_\gamma$, plotted versus the rupture-zone size $L_r$. The model parameters are $a=1$, $b=9$, $c=1000$, $l=20$, $v^*=10^{-2}$ and $\nu=10^{-8}$. (b) The mean duration times averaged over all events plotted versus the friction parameter $b$ with $a=1$ (main panel), and versus the friction parameter $a$ with $b=9$ (inset).  The other parameters are $c=1000$, $l=20$, $v^*=10^{-2}$ and $\nu=10^{-8}$. (c) The mean duration times plotted versus the crossover velocity $v^*$ with $l=20$ (main panel), and versus the inverse stiffness parameter $1/l$ with $v^*=10^{-2}$ (inset). The other parameters are $a=1$, $b=9$, $c=1000$, $l=20$ and $\nu=10^{-8}$.
}
\end{figure}

\subsection{C. The continuum limit}

 In view of the intrinsic discreteness of the BK model, it would be important to clarify the fate of the nucleation process in its continuum limit. We have shown above that the condition of whether the block size, an intrinsic short-length cutoff scale of the model, is larger or smaller than the nucleation length $L_{sc}$ largely affects the nature of the nucleation process. In particular, the continuum limit of the BK model always lies in the weak frictional instability regime. This gives us an important suggestion that an earthquake at a mature homogeneous fault obeying the RSF law always accompanies the quasi-static nucleation process \cite{Rice}.

 As mentioned, the continuum limit of the BK model corresponds to making the block size to be infinitesimally small $d\rightarrow 0$, simultaneously making the system infinitely rigid $l\rightarrow \infty$ so that $d=1/l$ \cite{MoriKawamura08c}. The equation of motion in the continuum limit has been given in the dimensionful form by eq.(\ref{continuumeq}). It should be emphasized that the length unit scaling the block size is $v_s/\omega$, while the length unit scaling the block displacement is the characteristic slip distance ${\mathcal L}$. Note that the former length scale, $v_s/\omega$, is absent in the standard continuum elasto-dynamic equation. The appearance of such a second length scale, in addition to the length scale of the critical slip distance ${\mathcal L}$, has occurred in the present model due to the existence of the characteristic time scale $\omega^{-1}$ borne by the $-\omega^2 U$ term in eq.(\ref{continuumeq}), which represents the plate drive directly applied to the fault layer as modeled by the block assembly of the BK model. 

 Let us examine the continuum limit of the two types nucleation lengths, $L_{sc}$ and $L_c$. Let us begin with $L_{sc}$. The continuum limit of $L_{sc}$ in the dimensionless form is given by $\tilde L_{sc}=\lim_{d\rightarrow 0} L_{sc}d=\lim_{l\rightarrow \infty} L_{sc}/l$. From the obtained analytical expression of $L_{sc}$, eq.(\ref{Lsc}), one can easily get
\begin{equation}
\tilde L_{sc} = \frac{\pi}{\sqrt{b-1}}.
\label{Lsc2}
\end{equation}
Remembering that the length unit here is $v_s/\omega$ and $b=B {\mathcal N}/(k_p\mathcal{L})$ (${\mathcal N}$ is the normal load), one can derive  the expression of the dimensionfull nucleation length in the continuum limit, $L_{sc}^\times$, as
\begin{eqnarray}
L_{sc}^\times &=& \frac{\pi}{\sqrt{\frac{\sigma_n v_s}{G\omega {\mathcal L}}B-1}} \frac{v_s}{\omega} \\ 
   &\simeq& \pi \sqrt{\frac{Gv_s{\mathcal L}}{\sigma_n \omega B}} , \ \ \ {\rm for}\ b>>1. 
\label{Lsc3}
\end{eqnarray}
Among the frictional parameters, $B$, not $B-A$, enters into the formula above. This is consistent with the earlier observation by Dieterich \cite{Dieterich92}, who derived the expression of the nucleation length dependent only on $B$, eq.(\ref{nucleationlength2}),  on the assumption of $v\theta >> 1$, which is also the condition we observed here.

 The derived expression of $L_{sc}^\times$ is a decreasing function of the frictional parameter $B$ and the normal stress $\sigma_n$, and an increasing function of the characteristic slip distance $\mathcal{L}$ and the rigidity $G$. This tendency is qualitatively consistent with the one indicated by the standard form, eq.(\ref{nucleationlength2}). However, the present formula of $L_{sc}^\times$ is different from eq.(\ref{nucleationlength2}) in that $L_{sc}^\times$ is inversely proportional to the square root of $G\mathcal{L}/(\sigma_nB)$, not to $G\mathcal{L}/(\sigma_nB)$ itself as in eq.(\ref{nucleationlength2}), the remaining part being complemented by the square root of the second length scale $v_s/\omega$. This difference originates from the difference in the expression of the stiffness $K$, eq.(\ref{stiffness}), versus the standard form in the continuum of $K\propto 1/L$. As mentioned, this difference can further be traced backed to the existence of the two length scales in the BK model, {\it i.e.\/}, the critical slip distance ${\mathcal L}$ and the length scale $v_s/\omega$, in contrast to only one length scale ${\mathcal L}$ in the standard elasto-dynamic model.

 Concerning the continuum limit of $L_c$, since the ratio $r=L_c/L_{sc}$ turns out to be hardly dependent on $l$ in the weak frictional instability regime relevant to the continuum limit, the dimensionful nucleation length in the continuum limit $L_c^\times$ is given by
\begin{equation}
L_c^\times = r(b) L_{sc}^\times\simeq (0.1b+4.4)L_{sc}^\times, \ \ \ b=\frac{\sigma_n v_s}{G \omega{\mathcal L}}B ,
\label{Lc}
\end{equation}
where $b$ is a number characterizing the fault interface.

 We also examine the continuum limit of the duration times of the nucleation process, $T_\alpha$, $T_\beta$, $T_\beta^\prime$ and $T_\gamma$. As shown in Fig.18(c), the $1/l$-dependence of these duration times turns out to be rather weak. This means that the duration times in the continuum limit should be close to the ones computed here for the discrete model.

\section{VI. Summary and discussion}

 We studied the nature of the nucleation process of the BK model in one dimension obeying the RSF law. The model turned out to exhibit qualitatively different nucleation phenomena depending on whether the frictional instability is either ``strong'' or ``week''. The condition of the strong or the weak frictional instability is simply given by $b>b_c$ or $b<b_c$, respectively, with $b_c=2l^2+1$. The quasi-static nucleation process, {\it i.e.\/}, the initial phase, exists only for the weak frictional instability. Two kinds of nucleation lengths, $L_{sc}$ separating the initial and the acceleration phases, and $L_c$ separating the acceleration and the high-speed rupture phases, were identified. The nucleation length $L_{sc}$ and the initial phase exist only in the weak frictional instability regime, while $L_c$ and the acceleration phase exist for the both regimes. The analytic expression of $L_{sc}$ was obtained as in eq.(\ref{Lsc}), which took the form of eqs.(\ref{Lsc2}) and (\ref{Lsc3}) in the continuum limit, while that of $L_c$ in the continuum limit was obtained as in eq.(\ref{Lc}). In fact, both $L_{sc}$ and $L_c$ were determined by the material parameters only, independent of the size of the ensuing mainshock. It means that {\it the information on $L_{sc}$ or $L_c$ cannot used for predicting the size of the subsequent mainshock\/}. Since the continuum limit of the BK model lies in the weak frictional instability regime, {\it an earthquake at a mature homogeneous fault under the RSF law always accompanies the quasi-static nucleation process\/}. When the discreteness or the inhomogeneity is strong, by contrast, an earthquake does not accompany the quasi-static nucleation process.

 Throughout the initial phase up to $L_{sc}$, the block sliding is extremely slow of order the loading speed of the plate. Beyond $L_{sc}$, the system gets into the irreversible acceleration phase where both the block sliding and the rupture propagation accelerate rapidly. Two characteristic points are identified within the acceleration phase. One is the point $v\simeq v^*$ where the block sliding velocity exceeds the friction crossover velocity, beyond which the rupture propagation is changed from the exponential to the super-exponential growth. The other is the point  $v\simeq v_{inertia}$ where the inertia effect becomes relevant, beyond which the block acceleration tends to be suppressed at the epicenter block due to the inertia effect. At $L\simeq L_c$, the sliding velocity $v$  of the epicenter block reaches its maximum, while the state variable $\theta$  of the epicenter block reaches its minimum. Beyond $L=L_c$, the epicenter block rapidly decelerates and stops. The system then gets into the high-speed rupture of a mainshock where the rupture front propagates in both direction with a nearly constant speed of the wave velocity. In the case of the strong frictional instability, a characteristic oscillatory behavior takes place at an early stage of the high-speed rupture, which is caused by multiple reflections of the rupture front.
 
 Various duration times of each stage of the nucleation process were studied. {\it The duration times also have no pronounced correlation with the size of the ensuing mainshock\/}. Particular attention was paid to the duration time of the acceleration phase $T_\beta$ and the remaining time available for a mainshock forecast $T_\beta^\prime$. Both $T_\beta$ and $T_\beta^\prime$ hardly depend on the model parameters, with the exception of the friction crossover velocity $v^*$, which tends to increase with decreasing $v^*$.  We argue that the remaining time for an earthquake forecast could be one or two magnitudes longer than the duration time of a mainshock, but perhaps not much longer than that.

 Next, with our present findings on the BK model in mind, we wish to discuss possible implications of the results to the nucleation process of real seismicity.  Of course, since the reliability of the 1D BK model in connection with real seismicity may be limited at the quantitative level, such implications to real seismicity should be taken only as indications. 

 Let us estimate the typical scales of these nucleation lengths on the basis of eqs.(\ref{Lsc3}) and (\ref{Lc}). Concerning $L_{sc}$, if we substitute the parameter values $\mathcal{L} \simeq 1$ [cm], $B\simeq 10^{-2}$ and $\frac{v_s}{\omega} \simeq 2$ [km] into eq.(\ref{Lsc3}), we get $L_{sc}^\times$ several kilometers.  Even though our expression of $L_{sc}$, eq.(\ref{Lsc3}), is different from the standard one, this value is not much different from, perhaps slightly greater than the corresponding estimates reported in the literature based on eqs.(\ref{nucleationlength}) and (\ref{nucleationlength2}). Concerning $L_c$, if we substitute typical parameter values in eq.(\ref{Lc}), $L_c^\times$ would be around 10 [km].

 Any possibility of an earthquake forecast lies in the acceleration phase. The remaining time $T_\beta^\prime$ plays an especially important role here. Let us estimate various duration times on the basis of our present results. If we revive the normalization units and substitute the typical parameter values, we get, for $v^*=10^{-4}$, $T_\alpha\simeq 10^2$ [year], $T_\beta\simeq 1$ [day], $T_\beta^\prime\simeq 1$ [hour] and $T_\gamma\simeq 1\sim 2$ [min]. For smaller $v^*$, $T_\beta^{\prime}$ could be even longer. However, as can be seen from Fig.18(c), the increase of $T_\beta^{\prime}$ with decreasing $v^*$ tends to be suppressed and to saturate for $v^*\lesssim 10^{-4}$. Hence, we deduce that, irrespective of the detailed value of the friction crossover velocity $v^*$, {\it the remaining time available for a mainshock forecast would not be much longer than several hours\/}. Hence, the time left seems not so long even in the best condition.

 The duration times $T_\beta$ and $T_\beta^\prime$ turn out to depend on the friction parameter $v^*$. Th friction crossover velocity $v^*$ is introduced in our analysis to describe the state at rest phenomenologically. In view of such a slow speed of the plate drive $\nu\simeq 1 $ [nm/sec], being of ``atomic'' scale, the question of whether the stuck region of the fault is completely stuck with a zero sliding velocity, or it is moving with a speed much lower than $\nu$, sounds too ``academic''. In  describing a macroscopic earthquake phenomenon, it would perhaps be more realistic to regard the stuck state as being completely at rest with $v=0$, and modify the relevant friction law so that it can describe the state at rest. Remember that the standard $a$-term proportional to $\ln v$ gives an infinitely negative friction for $v\rightarrow 0$, and does not allow anything to stop whatsoever. In other words, we feel that considering the ``stuck'' state as a state with its sliding velocity $0<v<<\nu$ is not much meaningful. Then, in order to describe such a state at complete rest $v=0$, we need a modified $a$-term with a nonzero crossover velocity $v^*$ ($ > \nu$), as was done phenomenologically here.

% To measure this friction crossover velocity $v^*$ discriminating the state at rest and the state obeying the logarithmic behavior from rock friction experiments, one should probably perform non-stationary-type measurements where the rock interface initially at rest begins to move, or its inverse, {\it i.e.\/}, taht initially moving comes to a complete stop, rather than the stationary-type measurements where the rock interface is driven with a constant sliding speed.    

%eoretical analysis of the RSF law regarding the stick-slikp process as a thermal-activation process certainly yields the form assumed here, but the expression of the friction crossover velocity $v^*$ contains empirically unknown `microscopic ' parameters. This uncertainty originates from the lack of our knowoedge that what actually bears the thermal-activation-type friction process `microscopically', dislocation, grain boundary, some group of molecules... ? As such, we have no reliable information about $v^*$ from theory at the moment. Yet, since recent experiments have constantly reported the pure logarithmic friction law for the $a$-term, the $v^*$-value would be at least pretty small. One should remember that the determination of each of the friction law is necessarily indirect even experimentally so that there always remains some uncertainty in determining the form of the friction law and its parameter values accurately.

 To predict the size of an earthquake would be even more difficult. Any quantity related to the nucleation process studied here, including the nucleation lengths $L_{sc}$ and $L_c$ and various duration times of the nucleation process, has no pronounced correlation with the size of an ensuing mainshock, at least for larger ones. {\it The problem of how big a mainshock is going to be is related to the stress state of the entire area, not limited to the nucleus area\/}. Just the information of the nucleus area is not enough to predict the ensuing mainshock size. If so, a wide-area survey of the stress state would be necessary for the detection of the mainshock size.

 Finally, we wish to discuss possible extensions of our present analysis. First, as the present model is one-dimensional, an obvious extension is to study the properties of the corresponding two-dimensional model. In two dimensions, the geometry could be more complex than in one dimension, which might modify at least a part of the results obtained here for the one-dimensional model.

 Second, in the present model, the nearest-neighbor interaction has been assumed  between blocks. In real earthquake faults, the existence of the crust perpendicular to the fault plane mediates the long-range interaction even between blocks away on the fault plane. In fact, the long-range interaction has been employed in the elastic-continuum analysis \cite{Dieterich92,RubinAmpuero,AmpueroRubin}. Even within the discrete BK model, the effects of the elastic long-range interaction was investigated, mainly concerning with its statistical properties such as the magnitude distribution \cite{MoriKawamura08b}. It would be desirable to study the nature of the nucleation process of such a long-range BK model, and compare it with that of the short-range model studied here.

 Third, the present model is homogeneous except for its intrinsic discreteness in the form of blocks. Real faults are more inhomogeneous where the elastic and the frictional parameters exhibit inhomogeneous distribution. The form of such a spatial inhomogeneity might be either random or more organized as being hierarchical \cite{Ide}. Within the BK model, it is possible to take account of such an inhomogeneity by assuming the model parameters varying from block to block \cite{CaoAki}.

 Fourth, the effects of the viscosity or the relaxation were not taken into account in the present model. Such relaxation effects should more or less exist in real faults.  It would also be desirable to clarify its role not only in the earthquake nucleation process but also in the mainshock itself. We leave these extensions and open problems to a future task.

In summary, we studied the properties of the earthquake nucleation process of a mature fault both  numerically and analytically on the basis of the spring-block BK model obeying the RSF law. We find that this simplified model successfully reproduces various features of the expected earthquake nucleation process. We analyzed the dynamical properties of the model at each stage of the nucleation process in detail, including their continuum limits, and further discussed the connection to a possible earthquake forecast.

\begin{acknowledgments}
The authors are thankful to T. Okubo, N. Hatano, N. Kato, T. Uchide and N. Ito for useful discussion. This study was supported by Grant-in-Aid for Scientific Research on Priority Areas 19052006. We thank ISSP, Tokyo University for providing us with the CPU time.
\end{acknowledgments}

\end{document}